\documentclass[sigconf,balance=false,dvipsnames]{acmart}
\usepackage{popets}

\setcopyright{popets}
\copyrightyear{YYYY}

\acmYear{YYYY}
\acmVolume{YYYY}
\acmNumber{X}
\acmDOI{XXXXXXX.XXXXXXX}
\acmISBN{}
\acmConference{Proceedings on Privacy Enhancing Technologies}
\usepackage{amsmath}
\usepackage{booktabs}

\usepackage{colortbl}
\usepackage{diagbox}
\usepackage{enumitem}
\usepackage{makecell}
\usepackage{multirow}
\usepackage{subcaption}
\usepackage{svg}
\usepackage[most]{tcolorbox}
\usepackage{threeparttable}
\usepackage{url}
\usepackage{xspace}

\begin{document}

\newcommand{\system}{\textsc{ProxyGPT}\xspace}
\newcommand{\repo}{{\urlstyle{tt}\url{https://github.com/dzungvpham/proxygpt}}}
\renewcommand\theadfont{}
\newcommand{\tbox}[1]{
    \begin{tcolorbox}[colback=gray!20, colframe=black, arc=3mm, boxrule=1pt, boxsep=0mm]
    #1
    \end{tcolorbox}
}

\title{\system: Enabling User Anonymity in LLM Chatbots via (Un)Trustworthy Volunteer Proxies}

\author{Dzung Pham, Jade Sheffey, Chau Minh Pham, Amir Houmansadr}
\affiliation{
    \institution{\texttt{\{\href{mailto:dzungpham@cs.umass.edu}{dzungpham}, \href{mailto:jsheffey@cs.umass.edu}{jsheffey}, \href{mailto:ctpham@cs.umass.edu}{ctpham}, \href{mailto:amir@cs.umass.edu}{amir}\}@cs.umass.edu}\\}
    \institution{\Large{University of Massachusetts Amherst, USA}}
  \country{}
}

\renewcommand{\shortauthors}{Pham et al.}

\NewDocumentCommand{\notation}{m o}{%
  \ensuremath{\mathcal{#1}%
  \IfValueT{#2}{_{#2}}}%
  \xspace%
}
\NewDocumentCommand{\user}{o}{\notation{U}[#1]}
\NewDocumentCommand{\proxy}{o}{\notation{P}[#1]}
\NewDocumentCommand{\chatbot}{o}{\notation{C}[#1]}
\newcommand{\ttp}{\notation{TP}}
\newcommand{\notary}{\notation{N}}
\newcommand{\broker}{\notation{B}}
\NewDocumentCommand{\adversary}{o}{\notation{A}[#1]}
\NewDocumentCommand{\query}{o}{\notation{Q}[#1]}
\NewDocumentCommand{\response}{o}{\notation{R}[#1]}
\NewDocumentCommand{\webproof}{o}{\notation{Z}[#1]}

\newcommand{\chau}[1]{{\color{teal}\{\textit{#1}\}$_{chau}$}}

\begin{abstract}
Popular large language model (LLM) chatbots such as ChatGPT and Claude require users to create an account with an email or a phone number before allowing full access to their services.
This practice ties users' personally identifiable information (PII) to their sensitive conversational data, thus posing significant privacy risks.
Unfortunately, existing private LLM solutions based on cryptography or trusted execution environments (TEEs) remain unpopular due to their prohibitive computational expense and platform restrictions.
To enable practical user anonymity in LLM chatbots, we propose \system, a privacy-enhancing system that leverages \emph{browser interaction proxies} to submit user queries on their behalf.
Unlike traditional proxy systems, \system operates at the ``user'' layer by proxying user interactions with the browser in identity-required environments, thus easily supporting a wide range of chatbot services.
We prevent malicious proxies by performing regular integrity audits using modern \emph{web proof} protocols for TLS data provenance.
We further utilize state-of-the-art LLM prompt guards on the proxy's side to mitigate unwanted user requests.
Additionally, we incorporate a give-and-take economy based on Chaum's blind-signature e-cash to incentivize \system users to proxy for others.
Our system evaluation and user study demonstrate the practicality of our approach, as each chat request only takes a few additional seconds on average to fully complete.
To the best of our knowledge, \system is the first comprehensive proxy-based solution for privacy-preserving AI chatbots.
\end{abstract}


\maketitle

\section{Introduction}

Chatbots such as ChatGPT, Claude, and Gemini~\cite{chatgpt, claude, gemini} have fundamentally transformed how people engage with artificial intelligence (AI) technologies.
Powered by large language models (LLMs) that are pre-trained on massive amounts of data, these chatbots can perform various complex tasks previously thought exclusive to humans, ranging from creative writing, programming, to medical, legal, or financial consulting~\cite{li_value_2024, rozière2024codellamaopenfoundation, clusmann_future_2023, li_large_2023, lai2023largelanguagemodelslaw}.
Despite their increasing popularity, LLM chatbots still \emph{require} users to share their personally identifiable information (PII) such as email addresses or phone numbers before enabling full access and functionality (Table \ref{tab:chatbots}).
At best, some chatbots (e.g. ChatGPT, Claude) may permit login-less access but with severely restricted features and rate-limited usage.
This practice of tying user identities to their conversations raises serious privacy concerns, especially given the sensitive nature of many interactions~\cite{mireshghallah2024trust}.
In addition to storing and reviewing user conversations~\cite{openai2024malicious, tamkin2024clio}, chatbot providers frequently train their LLMs on this data, potentially exposing private user information to LLM memorization and data extraction attacks~\cite{carlini2023quantifying, nasr2025extraction, neel2024privacyllmsurvey, miranda2025preserving}.

\begin{table}
    \setlength{\tabcolsep}{2pt}
    \normalsize
    \centering
    \begin{threeparttable}
        \caption{Identity requirement status by 10 popular LLM chatbots and 3 privacy-focused ones as of May 2025, with estimated monthly website traffic from \protect\url{semrush.com} (averaged over 3 months from January to March 2025). 3 out of the 10 non-private chatbots \emph{always} require users to log in, while the other 7 have a \emph{partial} identity requirement (login is necessary to access full functionalities).}
        \begin{tabular}{lcrr}
            \toprule
            \thead{Chatbot service} & \thead{Identity\\ required?} & \thead{Monthly\\ site visits} & \thead{Monthly\\ unique users} \\
            \midrule
            OpenAI ChatGPT~\cite{chatgpt} & Partial~\tnote{a} & 5.1B & 593.7M \\
            DeepSeek AI~\cite{deepseek} & \textbf{Always} & 366.7M & 93.2M \\
            Perplexity AI~\cite{perplexity} & Partial & 157.5M & 26.3M \\
            Google Gemini~\cite{gemini} & Partial~\tnote{a} & 137.3M & 76.6M \\
            Anthropic Claude~\cite{claude} & \textbf{Always} & 115.3M & 19.2M \\
            Microsoft Copilot~\cite{copilot} & Partial & 104.7M & 32.5M \\
            xAI Grok~\cite{grok} & Partial~\tnote{a} & 72.1M & 17.1M \\
            Quora Poe~\cite{poe} & \textbf{Always} & 46.4M & 9.2M \\
            Meta AI~\cite{meta_ai} & Partial & 11.7M & 3.7M \\
            Mistral AI Le Chat~\cite{lechat} & Partial & 11.2M & 2.5M \\
            \midrule
            DuckDuckGo DuckAI~\cite{duckai} & None & 1.4M & 0.4M \\
            AnonChatGPT~\cite{anonchatgpt} & None & 45.6K & 37.9K \\
            Brave Leo AI~\cite{leoai}~\tnote{b} & None & N/A & N/A \\
            \bottomrule
        \end{tabular}
        \begin{tablenotes}
            \footnotesize
            \item[a] Loginless mode does not currently support Tor.
            \item[b] Integrated into the Brave browser. No dedicated website available.
        \end{tablenotes}
        \label{tab:chatbots}
    \end{threeparttable}
\end{table}

We argue that users should be able to \emph{anonymously access} not only LLM chatbots but also AI technologies more broadly, for two key reasons.
First, anonymous interactions with chatbots and AI empower people to freely express their ideas and opinions without fear of judgment, surveillance, or retaliation, especially for marginalized groups~\cite{ashraf2020ai, sannon2022marginalized}.
Second, anonymous access to AI can promote user trust in this influential technology and further drive its adoption, thus potentially encouraging chatbot and AI providers to adopt a privacy-first development approach.
Motivated by these benefits of user anonymity, in this paper, we attempt to tackle the following research question:
\tbox{\textbf{RQ}: How can we design and implement a practical system to enable user anonymity in identity-required LLM chatbots?}

\textbf{Shortcomings of current approaches:}
Recent attempts at private LLM chat suffer from various technical and design limitations that hinder their widespread adoption.
One direction is based on \emph{cryptography} techniques, such as secure multiparty communication and homomorphic encryption, which let users perform private LLM inference while preventing the LLM owners from accessing the content of the conversations~\cite{tianyu2022thex, pang2024bolt, lu2025bumblebee, key2025shaft}.
This strong guarantee, however, comes at the expense of \textbf{prohibitive computational cost}.
For example, Bumblebee, one of the state-of-the-art SMC-based inference methods, takes eight minutes and several gigabytes of network overhead to generate a single token for the Llama 2 7B model, given only 8 input tokens~\cite{lu2025bumblebee}.
A more feasible direction is via \emph{trusted execution environments} (TEEs), which consist of specialized hardware and software~\cite{zhu2024tee}, with Apple's Private Cloud Compute and Meta's Private Processing for WhatsApp being one of the few proposed systems~\cite{applepcc, meta_private_processing}.
While faster, these industry-owned solutions not only \textbf{restrict users} to specific ecosystems and LLMs, but also require users to \textbf{trust in ``big tech'' companies}.
Furthermore, both cryptography and TEE require modifications to the chatbot services and cannot easily support all chatbot features (e.g., LLM-based web search).
A third direction is via chatbots such as DuckAI~\cite{duckai} and AnonChatGPT~\cite{anonchatgpt}, where user prompts are relayed through a single trusted party to a chatbot.
Although these VPN-like chatbots are more readily available than the other two solutions, they do not natively hide users' network-level identity and provide \textbf{no integrity guarantee} for the chatbots' responses.

\textbf{Our objectives:}
We explore alternatives that can simultaneously accommodate user anonymity and practicality.
In particular, we determine the following properties to be critical to the design of a privacy-preserving chatbot system:
\begin{enumerate}[left=2pt,label=\roman*.]
    \item \emph{Unlinkability}: Both network-level identities (e.g., IP addresses) and application-level identities (e.g., login credentials) of a user must not be easily linkable to their chat conversations.
    \item \emph{Multi-proxy}: Instead of relying on a single entity for privacy, the system should utilize multiple proxies who volunteer to help other users with privacy.
    \item \emph{Proxy verifiability}: Users should be able to formally verify the proxies' handling of their chat requests and the integrity of the corresponding chatbot responses.
    \item \emph{Abuse prevention}: Proxies should be able to filter out bad user requests that can violate the chatbots' terms of service.
    \item \emph{Ease of adoption}: The system should be easy to set up and use with well-established software stacks and also be inexpensive.
    \item \emph{Sustainability}: The system should incentivize participation while also mitigating Sybil \& Denial-of-service (DoS) attacks.
    \item \emph{Rich features}: In addition to responsive multi-turn conversations, the system should support a wide variety of chatbots.    
\end{enumerate}

\begin{figure}
    \centering
    \includegraphics[width=0.95\columnwidth]{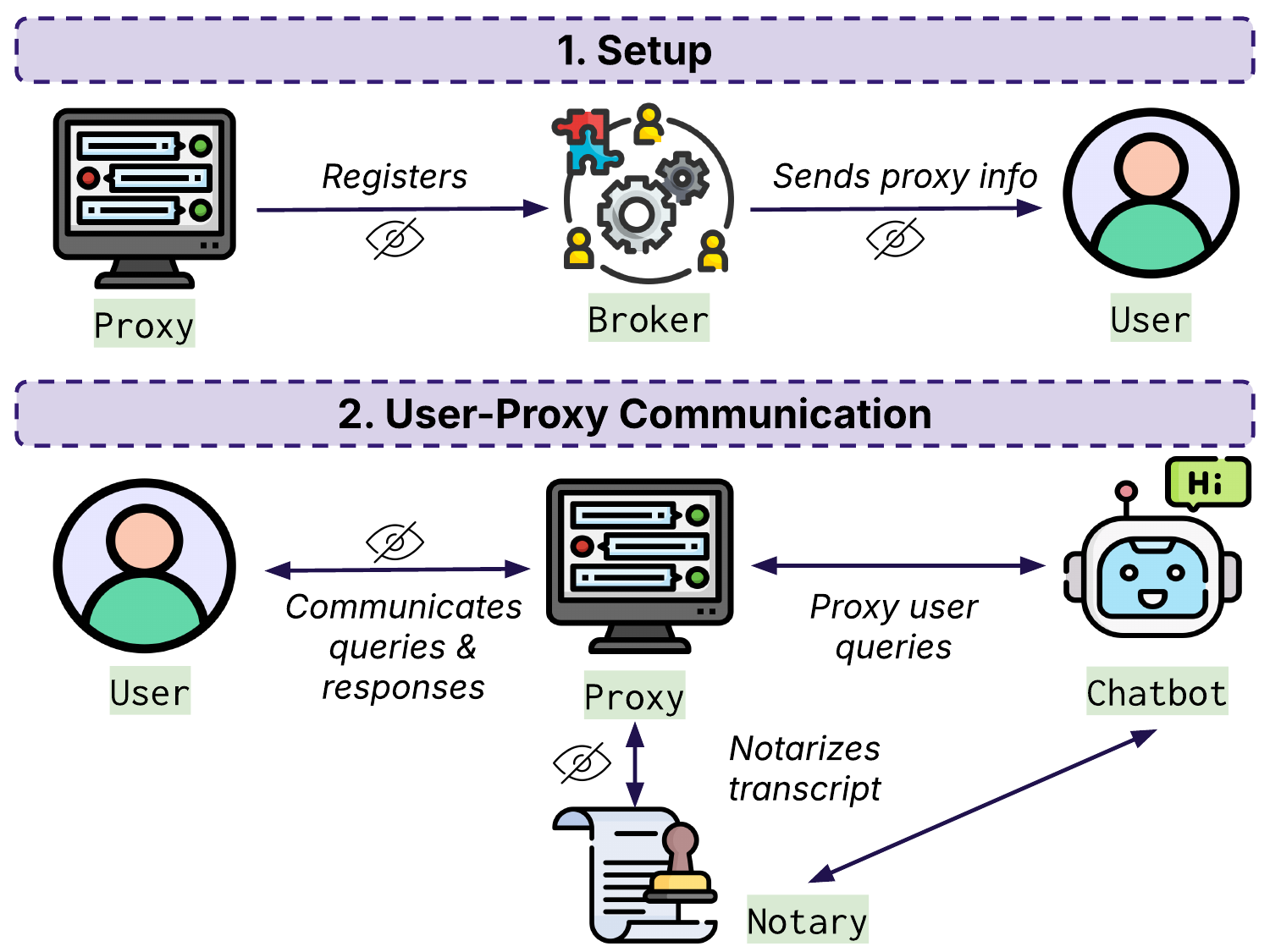}
    \caption{Overview of \system.
    Users select volunteer proxies via the broker.
    Proxies are (randomly) audited using a web proof protocol.
    All communications (except between proxies and chatbots) utilize AC protocols like Tor.}
    \label{fig:framework}
\end{figure}

\textbf{Our proposed solution:}
Achieving all of these objectives is a nontrivial task and requires balancing trade-offs carefully.
We propose \system, the first \textbf{peer-based} privacy-enhancing chatbot solution that can satisfy the aforementioned design goals (Figure \ref{fig:framework}).
By leveraging \emph{browser interaction proxies} who use their own chatbot accounts to interact with chatbots on behalf of anonymity-seeking users, the system can protect user identities at both the network and application levels from the chatbot providers.
User identities are also hidden from the proxies and the system using an anonymous communication (AC) protocol such as Tor~\cite{dingledine2004tor} or VPN~\cite{mullvad}.
To prevent proxies from tampering with the chatbot requests and responses, we perform \textbf{integrity audits} using modern \emph{web proof} protocols~\cite{zhang2020deco, celi2025distefano, kalka2024tlsnotary}, which can generate cryptographically verifiable proofs for the authenticity of the proxies' responses.
We also support \textbf{content moderation} by utilizing state-of-the-art prompt safety models~\cite{inan2023llamaguard} running in the browser to help proxies filter out undesirable user requests.
To keep \system \textbf{free to use}, we incentivize users to volunteer as proxies by rewarding them with lightweight \emph{electronic cash} (e-cash) based on Chaum's blind signature~\cite{chaum1990ecash}.
This e-cash can be obtained by proxying and completing audits, and can be spent on more \system conversations without linking the spender to the original e-cash obtainer.
The combination of integrity audits and e-cash also serves as a \textbf{DoS and Sybil mitigation} thanks to the computation cost involved.

We demonstrate the practicality of our design by implementing a \emph{fully functional} proof of concept (PoC).\footnote{Code available at \repo.}
Users can securely access \system through a modern website modeled after popular chatbot interfaces and hosted as a Tor hidden service (Figure \ref{fig:proxygpt_ui}).
The PoC offers support for \textbf{multiple popular chatbots}, specifically ChatGPT and Claude.
Proxies can participate simply by running a browser extension on any Chromium-based browser, and can optionally enable a browser-optimized Llama-Guard-3-1B~\cite{inan2023llamaguard} locally to exclude user requests they do not want to serve.
Our performance evaluation shows that \system adds $<$10 seconds on average to the full response latency, making it particularly suitable for time-consuming features such as ChatGPT's Deep Research~\cite{chatgpt_deepresearch}.
Our user experience study shows that despite the delay, \system offers a greater sense of privacy compared to traditional AI chatbots, particularly when users are hesitant to share their identities.

\section{Overview of \system} \label{sec:overview}

This section presents a high-level overview of \system, including its system, threat, and privacy models.

\subsection{System Model}

\subsubsection{Roles}
We consider four principal roles:
\begin{itemize}[left=2pt]
    \item User \user: Anyone who wishes to use chatbot services anonymously. \user can issue a chatbot conversation request, which we refer to as \query (for query).
    \item Chatbot service \chatbot: A LLM-powered chatbot service with only black-box, identity-required access. We also sometimes refer to the provider of a chatbot as \chatbot since the provider can own the chatbot (e.g., OpenAI, Google, Anthropic). There can also be more than one \chatbot for users to choose from.
    \item Proxy \proxy: A volunteer who interacts with \chatbot on behalf of \user to submit \query and retrieve a response \response from \chatbot.
    \item Trusted (third) parties \ttp: System components considered as trusted, e.g., notary \notary and communication bridge/broker \broker.
\end{itemize}

\subsubsection{Identity definition} \label{sec:identity_defn}
When we discuss the `identity' of a user \user, we refer to two identity types:
\begin{itemize}[left=2pt]
    \item Application-level: Email addresses, phone numbers, device fingerprints, and any other types of application-layer credentials that can be collected by \chatbot when \user visits it.
    \item Network-level: IP addresses, TLS/SSL handshake info, etc.
\end{itemize}

\system thus aims to protect this definition of identity.
Similar to many other privacy-preserving systems (e.g., Tor), \system does not protect against identity leakage in the content of the chatbot conversations, though we do provide some limited support for PII detection in user prompts (Section \ref{sec:proxygpt_ui}).

\subsubsection{Protocol overview}
At a high level, \system consists of the following operations (Figure \ref{fig:framework}):
\begin{enumerate}[left=2pt]
    \item Query submission: \user securely and anonymously sends \query to \proxy over an AC protocol (e.g., Tor), possibly facilitated by \broker.
    \item Chatbot interaction: \proxy uses their service credentials to interact with \chatbot and submit \query to \chatbot, getting \response in response.
    \item Response delivery: \proxy securely and anonymously sends \response to \user over an AC protocol, possibly facilitated by \broker.
    \item Notarization: A notary \notary can notarize the integrity of \query and \response by executing a web proof protocol (e.g., TLSNotary~\cite{tlsnotary}) with \proxy and \chatbot, producing a proof \webproof.
    \item Payment: Upon a successful verification, \proxy can be rewarded with a privacy-preserving form of payment such as cryptocurrency or Chaumian e-cash to ``buy'' more \system queries.
\end{enumerate}

\subsection{Threat Model} \label{sec:threat_model}

We assume a \textbf{malicious} adversary \adversary that may control up to $t < n$ proxies \proxy as well as a single chatbot \chatbot (but not all chatbots).
This realistically reflects the scenario where the chatbot itself is an adversary and can volunteer to proxy.
\adversary may also participate in \system as a user \user.
Given these assumptions, \adversary has the following capabilities:
\begin{itemize}[left=2pt]
    \item \adversary can identify any \proxy that interacts with \chatbot, e.g., by posing as \user to submit a watermarked \query that can be uniquely traced back to the responsible \proxy.
    \item \adversary can rate-limit or restrict access to \chatbot for any identified \proxy.
    \item \adversary can stealthily modify the response \response sent by \chatbot to an identified \proxy, e.g., by injecting subtle watermarks not visible to humans~\cite{kirchenbauer2023watermark} in the hope that \user who receives \response will ``reuse'' the watermark in their non-anonymous chat sessions, leaking their identity.
    \item \adversary can control a compromised \proxy to fabricate the response \response sent to \user or the \query sent to \chatbot (e.g., to cut costs or inject malicious content to leak \user's identity via cross-site scripting).
    \item \adversary can directly observe and analyze all conversations with \chatbot to create a profile of each author, possibly to perform automatic author attribution using ML-based stylometry techniques~\cite{narayanan2012author}.
\end{itemize}

\subsection{Privacy Model} 

\subsubsection{Goals} \label{sec:privacy_defn}
\system's primary objective is to enable \textbf{sender anonymity}~\cite{pfitzmann2001anonymity}, specifically the \emph{unlinkability} between \user's identity (as defined in Section \ref{sec:identity_defn}) and their query \query against the adversary \adversary.
One way to capture this notion is via the `degree of anonymity' metric based on normalized Shannon entropy~\cite{diaz2002anonymity, wagner2018metrics}:

Let \user[i] denote user $i$ of all $n$ users \emph{suspected} by \adversary. $n$ is thus the \textbf{anonymity set size}.
Let \query be a proxied query observed by \adversary and $P_{\query}(\user[i])$ be the probability assigned by \adversary to \query originating from \user[i], with $\sum_{i}^{n} P_{\query}(\user[i]) = 1$.
\query's degree of anonymity is defined as:
$$priv(\query) \coloneq \frac{H(P_{\query})}{\max(H(P_{\query}))} = \frac{-\sum_{i}^{n}P_{\query}(\user[i]) \log_2{P_{\query}(\user[i])}}{\log_2{n}}$$
This metric depends on the anonymity set size $n$ and $P_{\query}$, and has range $[0, 1]$, with 0 indicating \adversary can only randomly guess and 1 indicating complete confidence in one user.
\system aims to minimize \adversary's confidence mainly by increasing $n$ via both an application- and network-level separation of user identities from the queries.
Additional features such as proxy integrity, content moderation, usability, etc., are also crucial to attract users to the platform to enlarge the anonymity set.
Consequently, these seemingly non-privacy aspects are also important security and privacy requirements for the system.
Note that the anonymity set size is further influenced by \adversary's belief regarding the user base, since \adversary does not need to consider all possible chatbot users if it can determine confidently which users participate in \system.

\subsubsection{Non-goals}
Firstly, \system does not aim to hinder \adversary's development of $P_{\query}$ based on \query's content.
Without SMC or TEE, the only way that we can see to prevent $P_{\query}$ would be to adversarially modify \query, e.g., to abstract information or to obfuscate the style.
As this is beyond the scope of our work, we thus assume that users can construct \query with minimal self-identifying leakage using existing prompt anonymization or authorship obfuscation techniques (Section \ref{sec:content_privacy}).
The exact formulation of $P_{\query}$ is up to \adversary's modeling choice and requires considerable exploration.
A major obstacle is the lack of a comprehensive dataset of real user conversations for external researchers to reliably model this problem.

Secondly, \system does not (and cannot) hide the identity of the proxies from \adversary as they need to directly interact with the chatbot using their own service credentials.
Concealing proxy identities would likely require modifying the content of the request and response without the user's knowledge to remove watermarks, but this contradicts our proxy integrity requirement and is also not guaranteed to work.
However, other than \adversary, proxy identities are hidden from the users and the trusted parties.

Lastly, since we consider proxy integrity as a privacy requirement, we need to distinguish this notion from chatbot integrity.
Without cooperation from the chatbot providers, it is difficult to verify whether the providers' output is authentic or not.
Existing work on verifiable inference is still far from being practical~\cite{sun2024zkllm}.
Realistically, however, we expect \chatbot will be reasonably honest to protect its reputation (e.g., \response may contain subtle watermarks but should remain usable).


\section{Details of \system} \label{sec:operations}

\begin{figure}
    \centering
    \includegraphics[width=0.95\columnwidth]{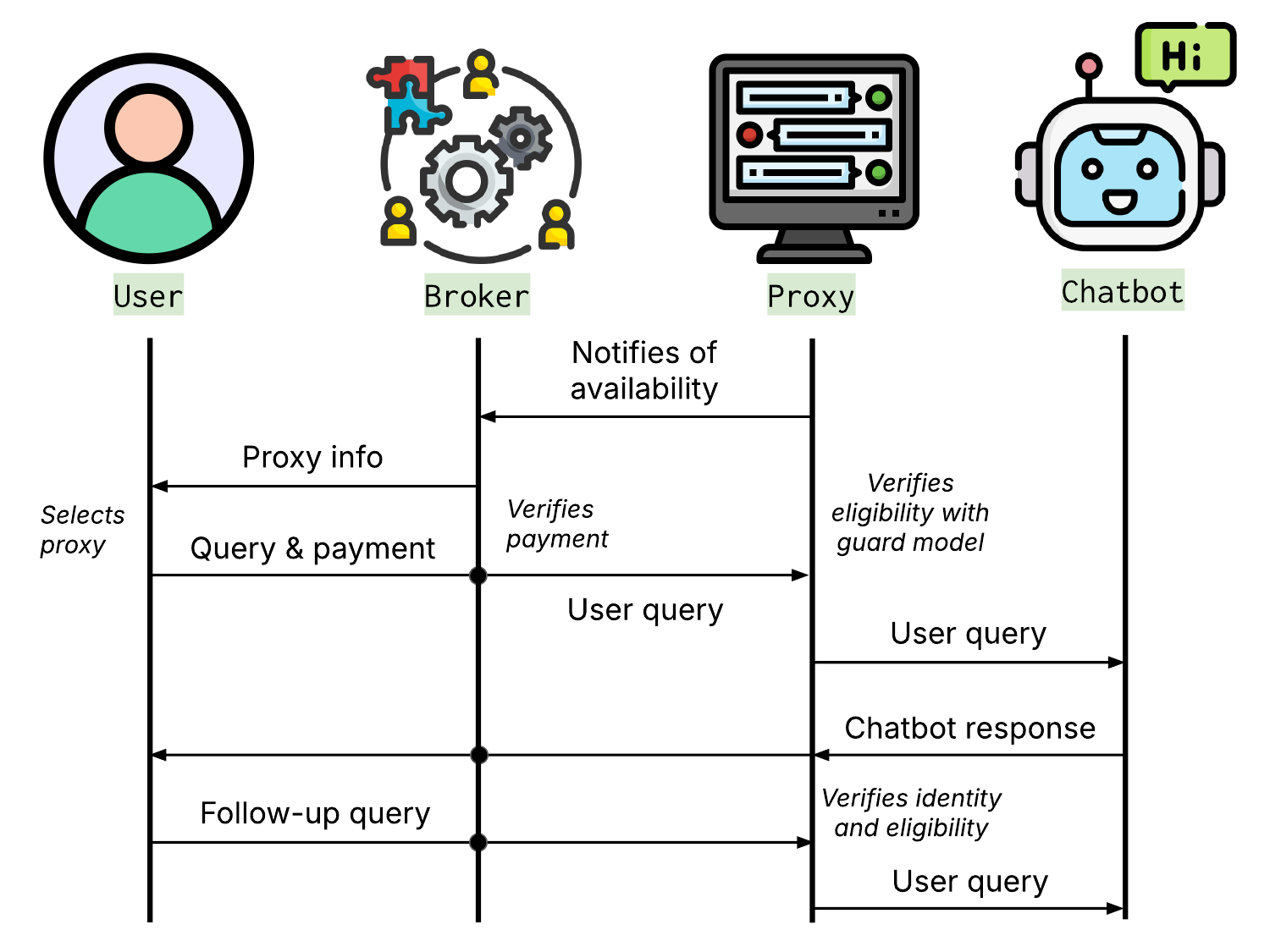}
    \caption{Illustration of query proxying process. The user sends their query to the proxy over the broker. The proxy checks the query's content (and identity if it is a follow-up to an existing chat thread) before sending it to the chatbot.}
    \label{fig:user_request}
\end{figure}

\begin{figure}
    \centering
    \includegraphics[width=0.95\columnwidth]{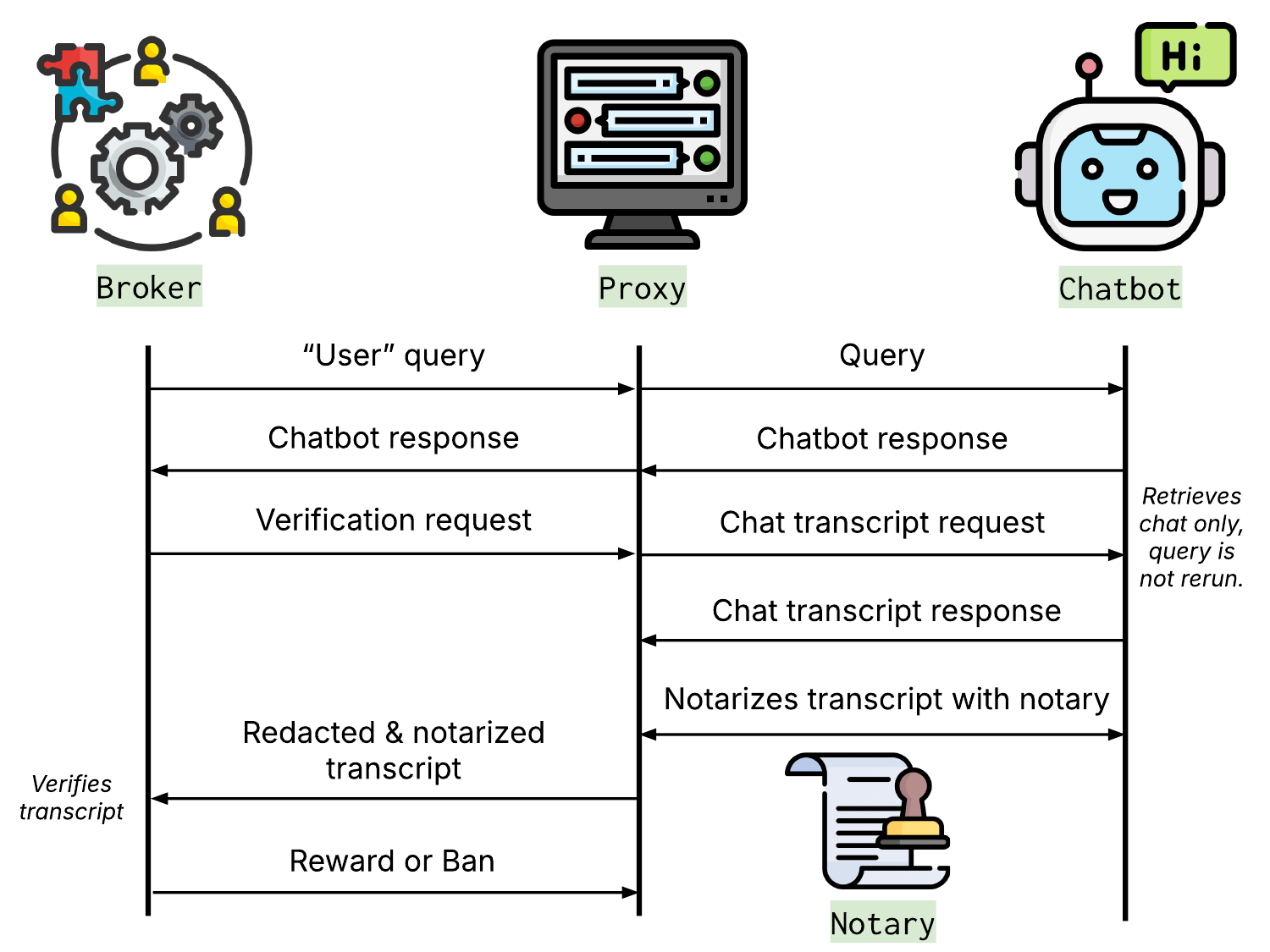}
    \caption{Proxy auditing. The broker sends the proxy a challenge query disguised as a normal one. After obtaining the proxy's response, the broker requests the proxy to participate in a web proof protocol like TLSNotary to ensure the original query and response are authentic.}
    \label{fig:audit}
\end{figure}

In this section, we describe in detail two important system components and two core operations in \system: submitting a query (Figure \ref{fig:user_request}) and verifying a proxy (Figure \ref{fig:audit}).

\subsection{Trusted System Components}

There are two trusted components in \system, namely the communication broker \broker and the notary \notary.
We keep their scope as narrow as possible; neither has direct access to user identities.
The use of these trusted services helps us avoid the complexity and cost of operating a fully decentralized system, which we intend to investigate as a follow-up work (Section \ref{sec:decentralization}).

\subsubsection{Broker}
The broker's main responsibilities include managing the directory of proxies, helping users discover proxies, and facilitating communication between users and proxies (similar to Tor's Directory Authority~\cite{dingledine2004tor}, Snowflake's central bridge~\cite{bocovich2024snowflake}, and MassBrowser's Operator~\cite{nasr2020massbrowser}).
Neither users nor proxies need to share their identities or the content of their chatbot queries with the broker.
Any new proxy must register with the broker and complete a verification process before it can participate in \system.
Users can contact the broker for a list of proxies and establish either a direct or a broker-facilitated communication channel with chosen proxies.
To ensure proxy integrity, the broker performs regular integrity audits with the help of the notary.

\subsubsection{Notary}
The notary's job is to verify and notarize the TLS-based communication session between a proxy and a chatbot service using a protocol like DECO~\cite{zhang2020deco} or TLSNotary~\cite{tlsnotary}.
The notarized session can then be independently verified by other parties such as the broker or the users themselves.
By design, the information visible to the notary/verifier only includes the server's hostname and any session data that has been selectively revealed by the proxies.
As such, the identities of proxies can remain hidden, while the text content of the chatbot conversation will need to be revealed for verification purposes.
We assume the notary does not collude with proxies to produce fraudulent notarization.

\subsection{Query Submission \& Processing}

\subsubsection{Info request}
A user who wants to use \system first anonymously requests a list of available proxies from the broker.
The broker retrieves its directory of proxies, filters for those that are recently active (e.g. last contact within the last five minutes), and then replies to the user with the proxies' contact information, such as their onion addresses, along with some performance statistics to help users choose.

\subsubsection{Proxy selection}

Users are provided with various proxy performance statistics calculated periodically by the broker to aid the proxy selection process.
These include mean response time, average daily request volume, and downvote rate (Section \ref{apd:stats}).

\subsubsection{Communication with proxy}

Using the selected proxy's contact information, users can establish a ``direct'' connection with the chosen proxy via an AC protocol like Tor.
If the connection succeeds, users can proceed to send their queries to the proxy (possibly with some form of payment).
Alternatively, if a direct connection is not possible as in the case of browser proxies, the broker can serve as a message dropbox for the users and queries.
This is our choice for the proof of concept because it does not require proxies to install any additional software beyond a browser extension (and Tor).
The query's content is end-to-end encrypted to prevent the broker from accessing.

\subsubsection{Query sanitization (optional)}

Before sending their requests to the proxy, users should ensure their queries do not contain any explicitly self-identifying information, particularly PII.
Users can use various PII detection tools to identify and redact/modify PII.
Users should also try to rewrite their queries to prevent authorship attribution.
We note that the techniques for text sanitization are still largely experimental and not guaranteed to work (Section \ref{sec:content_privacy}).

\subsubsection{Payment} \label{sec:payment}

Certain proxies may require users to pay for their services to offset the operational costs.
API usage in particular imposes a fee for every input and output token.
To preserve privacy, users can pay using cryptocurrencies with low transaction fees.
However, we recognize that requiring users to spend real money will likely hinder the adoption of \system.
As such, we design an alternative payment method where users can ``pay'' by volunteering to proxy for others.
For every successfully validated verification request, proxies can obtain from the broker a ``token'' that can be spent to buy more \system queries.
This payment model is especially suitable for browser-based proxies who can make free requests to chatbots via a web-based UI.
To prevent linkage of a proxy's token to their user queries, we use Chaum's blind-signature-based electronic cash~\cite{chaum1983blindsig, chaum1990ecash} to allow the proxy to privately spend their tokens without anyone learning the identity of the tokens' owner, even the broker.

Users should not use an e-cash token obtained from proxying for one chatbot to ask queries to the same chatbot, because the provider can then narrow down the list of suspected user identities to the pool of identified proxies.
The broker can enforce this isolation by issuing different types of e-cash for different chatbots and making sure that an e-cash cannot be spent on its corresponding chatbot.

\subsubsection{Rating proxy response (optional)}

Since a proxy may fabricate its response, users should ideally be able to request the proxy to provide a formal cryptographic proof of the response's authenticity via a protocol like DECO or TLSNotary.
However, due to the high overhead of these protocols, we limit the proof requirement to audit requests from the broker only.
Users can instead ``downvote'' bad responses, which will impact the proxies' reputation.

\subsection{Proxy Verification} \label{sec:proxy_verification}

\subsubsection{Registration}

A proxy must first register itself with the broker by completing an integrity audit before it can participate in \system.
The proxy initiates the registration by sending the broker a unique `pseudonym' for communication purposes (e.g., onion address in Tor or a public key), along with a list of constraints such as the supported chatbots.
The broker makes sure the proxy's pseudonym is unique, then responds with an integrity audit.
By requesting an audit right at the start, the broker can ensure that the proxy is capable of performing web proof protocols correctly.

\subsubsection{Audit preparation}
The broker issues an integrity audit to each proxy at the registration stage or randomly for each user request.
The audit involves a special chatbot query that the proxy must honestly complete within a reasonable time frame (e.g., 10 minutes).
For probabilistic audits, the content of the audit query should be difficult to distinguish from regular user queries (registration does not need this property since the proxy already knows it is being audited).

\subsubsection{Audit web proof}

After the broker obtains the proxy's response, it then requests the proxy to produce a web proof for its chat session.
The proxy performs a web proof protocol such as TLSNotary with the chatbot server and \system's notary, then sends the proof to the broker.
The proof should hide any sensitive information that may reveal the proxy's identity to the broker (Figure \ref{fig:tlsn_proof}).
The broker verifies the proxy's response and proof and notifies the proxy of the result.
If the verification fails, the proxy is disallowed from participating in \system.
Otherwise, the broker can reward the proxy, e.g., with a Chaumian e-cash.


\section{Implementation Details of Proof of Concept} \label{sec:implementation}

This section presents various implementation choices for our \system proof of concept.
We note that this represents only one possibility of how \system can be instantiated.

\subsection{Broker \& Notary}

\subsubsection{Deployment}
We deploy the broker as a Tor hidden service~\cite{dingledine2004tor} on an Ubuntu 20.04 virtual machine and implement it using the Flask web framework~\cite{flask}, Gunicorn Python WSGI server~\cite{gunicorn}, and Nginx as a reverse proxy~\cite{nginx}.
The broker exposes several REST endpoints for retrieving proxy info, registering proxies, sending/receiving queries and responses, etc.
It maintains a MySQL database that stores proxy information such as pseudonyms, registration status, performance metrics, and end-to-end-encrypted user-proxy communication data.
All communications are automatically deleted after 1 day.

For the notary, we deploy a Rust-based TLSNotary server (v0.1.0-alpha.5)~\cite{tlsnotary, kalka2024tlsnotary} in a separate Ubuntu 20.04 virtual machine along with a noVNC-based WebSocket proxy ~\cite{websockify}.
We choose TLSNotary because it is currently the only open-source, browser-friendly, and actively maintained web proof framework.
Unlike the broker, the notary is hosted publicly due to the prohibitive communication overhead when performing the TLS data provenance protocol over Tor.
As a result, proxies need to use a faster AC system, such as Mullvad VPN~\cite{mullvad}, to be able to execute TLSNotary while also protecting their privacy.
The WebSocket proxy is needed because browser-side code cannot open raw sockets required for TLSNotary.
(\system proxies can also use their own local WebSocket proxy instead of the publicly hosted one, but this requires an extra step.)

\subsubsection{Creating audit queries} \label{sec:audit}
As users send their queries to a proxy, the broker probabilistically inserts its own query disguised as a regular user request so that on average, each proxy would receive 4-6 fake queries before hitting the chatbot providers' hourly rate limit.
The broker also randomly requests web proof for the fake queries.
This \emph{data poisoning} strategy is meant to reduce the likelihood of the proxy's being able to distinguish between regular and audit queries based on the network/temporal pattern or the query content.
To further reinforce content indistinguishability, we randomly source the challenges from various question-answering platforms (e.g., Reddit, Quora, StackOverflow) and real user prompt datasets~\cite{zhao2024wildchat}.
We further rewrite the seeds to different topics while retaining their original styles using LLama-3.1-8B-Instruct\footnote{\url{https://huggingface.co/meta-llama/Llama-3.1-8B-Instruct}} to prevent proxies from simply checking if the seeds can be found online.
More complex techniques such as mixing different seeds together could also potentially be applied to confuse an adversary~\cite{pham2025frankentextstitchingrandomtext}.


\subsubsection{Verifying audit queries}
For verifying TLSNotary proofs, the broker hosts a local NodeJS server~\cite{nodejs} that runs the TLSNotary JavaScript API.
As the API is written for the browser environment, to enable it in NodeJS, we use a virtual DOM and ``monkey-patch'' the browser-specific Web Worker API with a cross-platform alternative to load local files.
After establishing the validity of the proof, the broker compares the original query and response with the selectively revealed text contained in the proof and checks the structure of the JSON response to ensure that there is no malicious manipulation (Figure \ref{fig:tlsn_proof}).

\subsection{User Interface} \label{sec:proxygpt_ui}

\subsubsection{Overview}
\begin{figure}
    \centering
    \includegraphics[width=\linewidth]{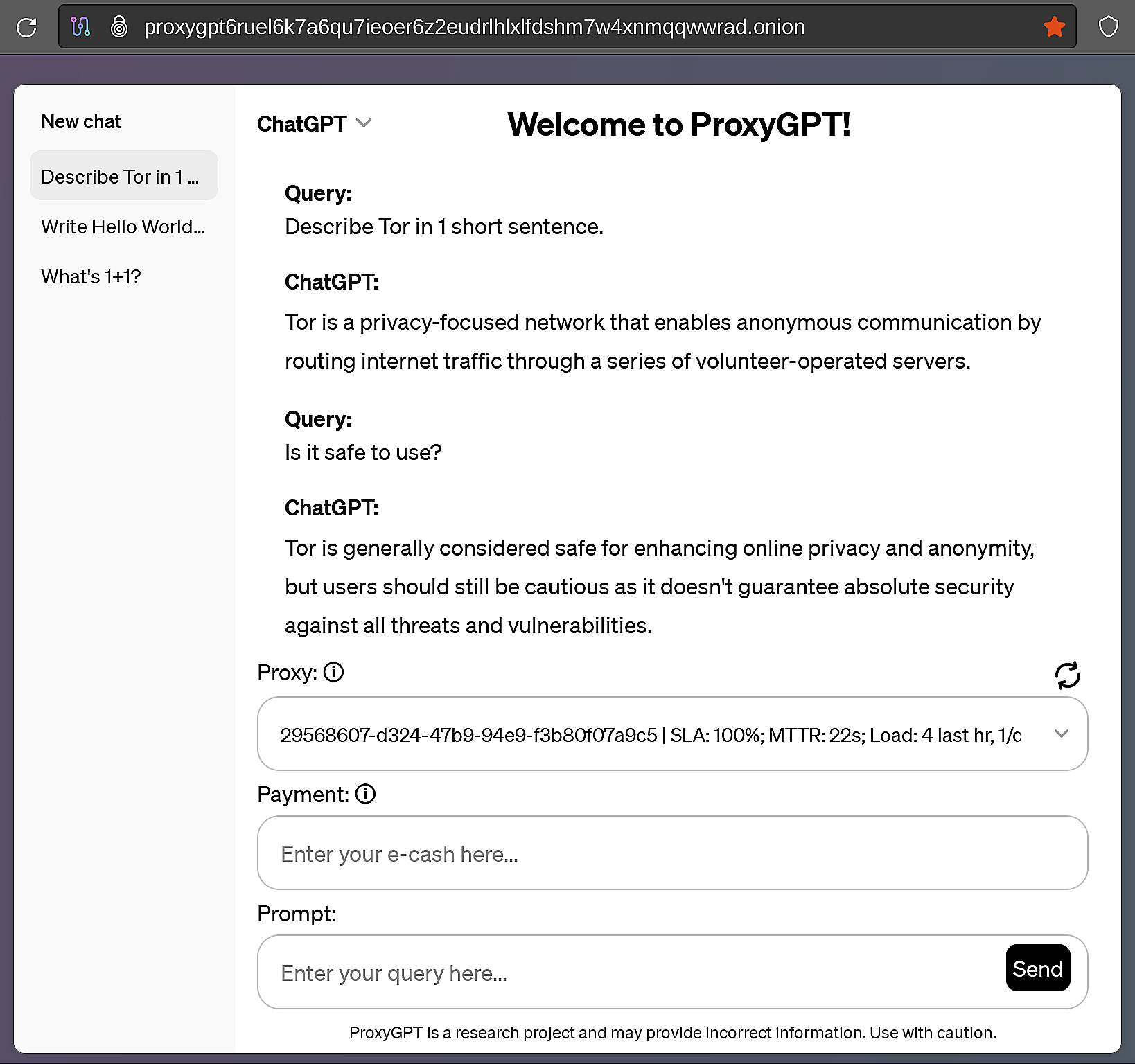}
    \caption{User interface of our \system website (screenshot taken from within the Tor browser). Users can switch between chatbots, select which proxy to use, engage in multi-query conversations, etc.}
    \label{fig:proxygpt_ui}
\end{figure}

The broker provides a simple and convenient Tor website for users to discover proxies and make their proxy requests (Figure \ref{fig:proxygpt_ui}).
The website's design is influenced by existing chatbots, particularly ChatGPT.
Users can easily switch between ChatGPT and Claude, engage in multi-query conversations, as well as downvote individual responses.
To preserve the styling and formatting of the original chatbot responses while also preventing cross-site scripting attacks (XSS), we use the same styling framework from ChatGPT and Claude and sanitize the responses with DOMPurify~\cite{dompurify}.

\subsubsection{Query encryption}
To submit a query in \system, the client-side logic generates an ECDH and an ECDSA key pair on curve P-256 for encrypting the query and decrypting the response end-to-end while also ensuring authenticity.
The client's payload is encrypted using an AES-GCM-256 key derived from the private ECDH key and the chosen proxy's public ECDH key.
The client keys are not reused across different queries to prevent them from being linked together.
In addition to the query content, extra options like which chatbot to use and which thread to continue are encrypted and thus not accessible even by the broker.
All cryptography operations use the browser-native WebCrypto API~\cite{w3cwebcryptoapi}.

\subsubsection{Query sanitization (optional)}
Users can optionally check their prompts for potential PII leakage using a browser-based PII detection model.
Specifically, we use the (half-precision) Piiranha-v1 model, one of the most popular PII detectors on HuggingFace,\footnote{\url{https://huggingface.co/iiiorg/piiranha-v1-detect-personal-information}} with the Transformers.js inference engine~\cite{transformersjs} and WebAssembly (WebGPU is not currently compatible with this model).
Note that downloading the >500MB model over Tor can take more than 5 minutes even on a 100Mbps Internet due to the onion routing's overhead.
However, users can modify Tor's settings manually to enable persistent caching to avoid having to re-download in subsequent sessions.

\subsection{Proxies}
\label{sec:browser_proxy}

\subsubsection{Overview}

\begin{figure}
    \centering
    \includegraphics[width=0.5\linewidth]{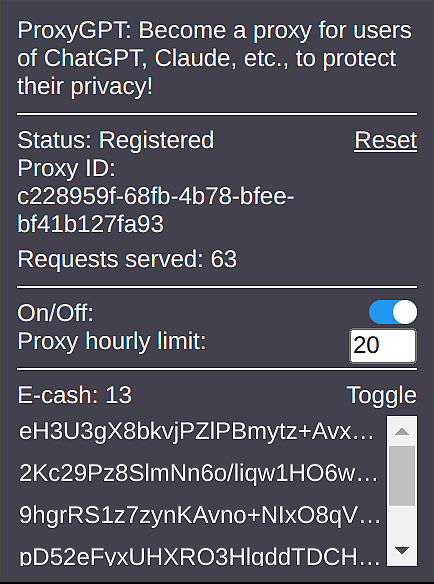}
    \caption{Pop-up page for \system's browser extension. It shows some basic statistics, allows users to easily turn the extension on/off and limit the number of proxy requests per hour, and displays the earned e-cash.}
    \label{fig:browser_extension}
\end{figure}

We implement a browser extension for Chromium-based browsers using the ManifestV3 extension standard~\cite{manifestv3} to enable users to become proxies with minimal setup (Figure \ref{fig:browser_extension}).
The extension programmatically interacts with the chatbot websites using the provided GUI components in the same way as regular users do.
Extension users must log in to the chatbot websites first and clear any bot detection mechanisms.
To prevent Chrome from pausing unfocused browser tabs and ensure a smooth proxying process without hindering normal chatbot usage, extension users should dedicate a separate browser application window to each chatbot website without including additional tabs.
Tor must also be installed and running in the background to enable communication between proxies and the broker (simply leaving the Tor browser open suffices).

\subsubsection{Registration}
Similar to the users, each proxy must generate an ECDH and ECDSA key pair, and then register with the broker using the generated keys.
To store the private keys on the browser side, we set the Web Crypto API to generate them with the extractibility feature disabled (i.e., not exportable to a non-binary format even with Javascript), then store the entire binary private key object using the IndexedDB API~\cite{pomcor2017keys}.
The public keys, on the other hand, are exported to the SPKI format and base64-encoded.

\subsubsection{Authentication}
Once registered, the proxy needs to authenticate itself prior to receiving queries by presenting a valid signature for a random nonce sent by the broker using the proxy's private ECDSA key.
Upon successful validation of the proxy's signature, the broker issues to the proxy a signed JSON Web Token (JWT)~\cite{rfc7519jwt} that confirms the proxy's authenticity.
The proxy can subsequently present the JWT to the broker to retrieve new queries and submit responses until the token's expiration.

\subsubsection{Handling queries}

The proxy processes each query one at a time to avoid overloading the chatbots with several queries simultaneously.
Each query is decrypted using the shared AES-GCM-256 key derived from the proxy's private ECDH key and the query owner's public ECDH key.
For queries that are follow-ups of existing chat threads, the proxy must validate the ownership by verifying a signature produced by the query owners of the IDs of the latest queries in the threads.
Note that the broker does not know whether a query is a new thread or part of an existing one since the payload is encrypted.
The volume of queries served can also be controlled by the proxy.

In order for the proxy to be able to support multiple different chatbot websites while also maintaining consistency with real user experiences, we use a unified GUI-based query submission flow consisting of the following steps:
\begin{enumerate}[label=\alph*.]
    \item Create/Select chat thread: For a new query, locate the `new chat' button/link. For a continuing query, locate the button/link for opening the chat thread with the relevant thread ID in the chat history. Click on the located element and wait for the website to load.
    \item Find input area and enter query: The input area is either a textarea or a div with the attribute ``contenteditable'' set to true. After inputting the query into the input area, dispatch any relevant input event to make sure the internal state of the website is updated correctly, then wait 1 second for the submit button to appear.
    \item Find submit button and click: Once the submit button appears and is usable, click on it and wait for a few seconds.
    \item Wait for response: Keep checking for DOM elements that indicate results are still streaming every second.
    \item Check for error: Chatbot providers may return server-side errors or enforce a rate limit.
    \item Retrieve query result: Once streaming is done, locate the DOM element corresponding to the latest message from the chatbot and extract the content (along with the generated thread ID in the updated URL).
\end{enumerate}

Once the response has been retrieved, the proxy encrypts it with the derived AES key above and sends it to the broker.
If any step above fails, the whole process is retried after a timeout of 1 minute.

\subsubsection{Handling audits}

\begin{figure}
    \centering
    \includegraphics[width=0.90\linewidth]{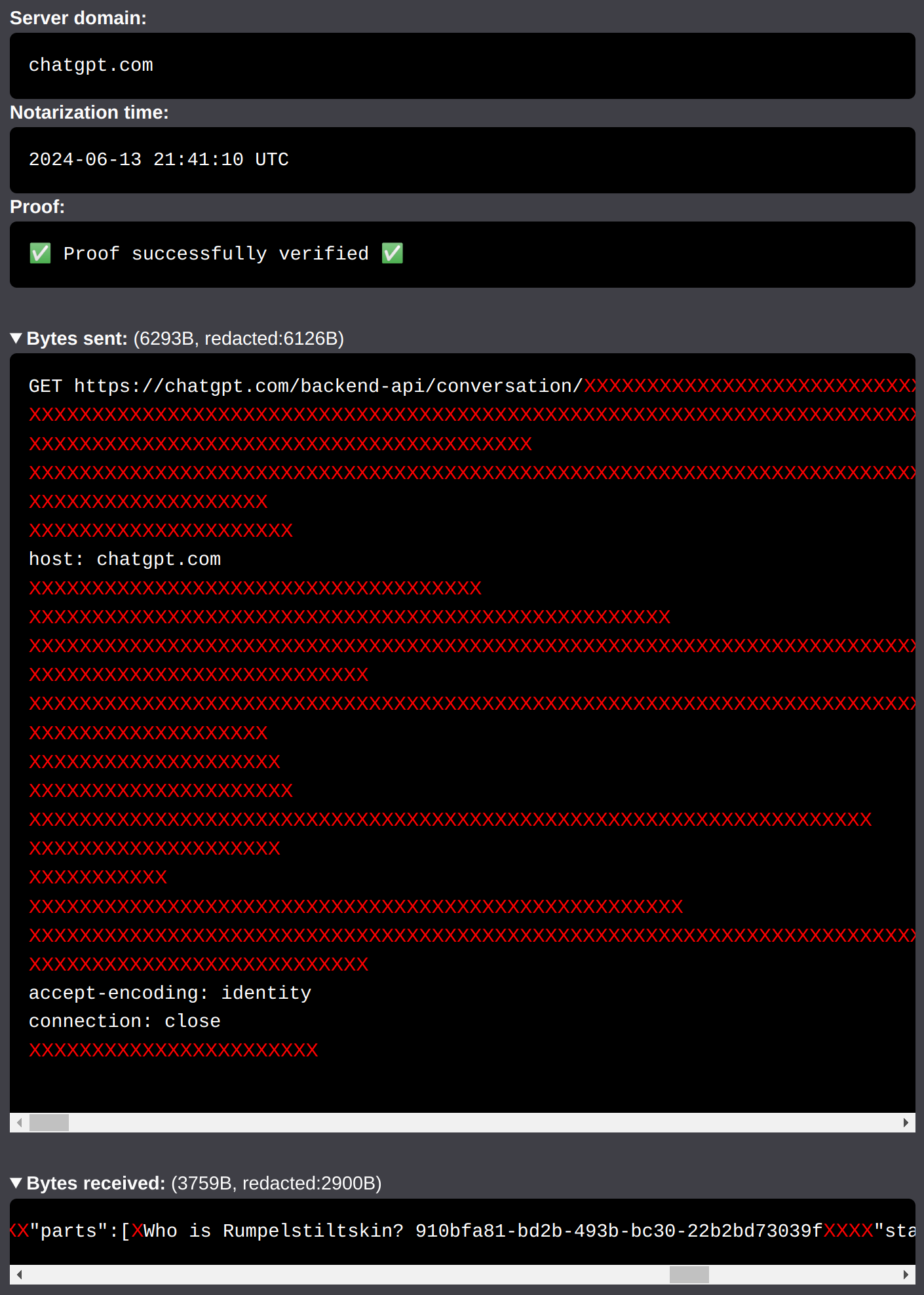}
    \caption{Visualization of a verified TLSNotary proof (screenshot taken from \url{https://tlsnotary.github.io/proof_viz/}). The notary can verify the chatbot's server domain and REST endpoint, the timestamp, and the content of the request. All other sensitive information, such as the authorization token, has been redacted by the proxy.}
    \label{fig:tlsn_proof}
\end{figure}

Proxies handle verification challenges using TLSNotary's WebAssembly-based JavaScript API (which we further customized to enable easier redaction of sensitive information in JSON format).
Proxies will participate in the three-way TLS protocol with our publicly hosted TLSNotary server, possibly over a VPN like Mullvad but not Tor due to the increased latency (once TLSNotary becomes sufficiently fast, Tor can be utilized to better protect proxy identity during audits).
To connect to the chatbot server, proxies can either use our public WS proxy or host their own local WS proxy like noVNC's Websockify~\cite{websockify} to better preserve their privacy.
Redacting sensitive information in the proof record involves hiding all HTTP request headers (the Request-URI in the Request-Line also needs to be partially hidden to avoid leaking the conversation ID) as well as parsing the JSON response to reveal only the relevant JSON structure and the content of the chatbot conversation (Figure \ref{fig:tlsn_proof}).
During this audit stage, proxies will not receive any new user requests.

\subsubsection{Query safety check (Optional)} \label{sec:prompt_safety_impl}

Proxies can optionally run Llama-Guard-3-1B~\cite{inan2023llamaguard} in the browser extension to filter out any unwanted user requests based on the content.
We compile and quantize the official model checkpoint\footnote{\url{https://huggingface.co/meta-llama/Llama-Guard-3-1B}} on HuggingFace to a browser-optimized MLC~\cite{mlc-llm} format with 4-bit weights and 16-bit activations (q4f16), reducing the model's size from 3GB to less than 850MB.
Proxies can perform inference efficiently using the WebLLM inference engine for browsers~\cite{ruan2024webllm}.
Note that this feature currently only works well for certain desktop browsers (e.g., Chrome, Microsoft Edge) with WebGPU~\cite{webgpu} and half-precision floating point (fp16) support (enabled on Windows and macOS but not on Linux as of this writing).
An alternative is to use an ONNX version of the model with the Transformers.js framework~\cite{transformersjs}, but based on our benchmark, this option is not as efficient, not to mention the larger (q4f16) model size.
By default, we use the official prompt guard template with 13 pre-defined unsafe prompt categories,\footnote{\url{https://www.llama.com/docs/model-cards-and-prompt-formats/llama-guard-3/}} but proxies can easily customize these to their own preferences.

\subsubsection{Obtaining e-cash}
Proxies can obtain a \system e-cash from the broker by successfully completing integrity audits.
We implement an e-cash library based on Chaum's RSA-based blind signature scheme~\cite{chaum1983blindsig, chaum1990ecash} using NodeJS's native `crypto' library with guidance from RFC 9474~\cite{rfc9474blindsig}.
To obtain e-cash, proxies first generate an appropriate random message, \emph{blind} it using the broker's e-cash public key, then attach the blinded result to the audit responses.
If the audit passes, the broker signs the blinded message using their private e-cash key and sends the signature back to the proxy.
The proxies can now \emph{unblind} the broker's signature to obtain a different signature that is valid but unknown to the broker.
The resulting signature and the generated random message together form the e-cash (Figure \ref{fig:browser_extension}).
This scheme thus allows \system e-cash to be spent without revealing the identities of the original owners.
When a coin is presented as payment, the broker must validate and verify that the coin has not been used.

\subsubsection{Proxy statistics} \label{apd:stats}
We present the following (hourly aggregated) statistics to users to help them choose proxies:
\begin{itemize}
    \item Service Level Agreement (SLA) compliance rate: We define the SLA as a proxy finishing a user request in less than 1 minute. The rate of SLA compliance can let users know how likely they will get a response within 1 minute.
    \item Mean time to respond (MTTR): MTTR is defined as the time it takes for a proxy to respond, averaged over all finished queries. Unprocessed queries are not included.
    \item Load: We calculate the average volume of requests sent to a proxy per day as well as the volume in the last hour. This shows how active or busy a proxy is.
    \item Downvote rate: Users can report or \emph{downvote} bad responses to help other users avoid low-quality proxies. This simple mechanism complements our more expensive cryptographic proxy verification technique.
\end{itemize}

\section{System Evaluation} \label{sec:evaluation}

In this section, we evaluate the performance of our \system proof of concept, focusing on the latency of its various operations. Our system only makes a small number of network requests with minimal payloads, so we do not report this information.
The only source of significant network overhead is when a proxy performs the TLSNotary protocol with the notary~\cite{tlsnotary, kalka2024tlsnotary}.
\system also only uses a sizable amount of storage and memory when the optional PII detection or prompt safety feature is enabled.

\subsection{Query latency}

We simulate a chat session by sending 30 different normal chat queries\footnote{This number is sufficient given our small margins of error calculated with a 95\% t-distribution confidence interval (Tables \ref{tab:perf_time} and \ref{tab:audit_time})} via our \system website to a ChatGPT proxy in the Google Chrome browser on an Ubuntu 20.04 machine with an Intel Xeon W-2225 CPU and 16GB of memory.
Each query has the format ``Tell me about $<$subject$>$ in (one $\mid$ two $\mid$ three) paragraph(s)'', where the subjects are chosen from a list of countries in the world, and the number of paragraphs is to induce responses of varying lengths (roughly between ASCII 800-1600 characters).
We measure the time taken for the following:
\begin{itemize}
    \item User delivering query to proxy: Involves the user submitting a request to the broker, who verifies the chosen proxy's identity and sends the query to the proxy.
    \item Proxy interacting with ChatGPT: Involves the proxy interacting with ChatGPT's website and notifying the extension's backend of the query results.
    \item Proxy delivering response to user: Involves the proxy sending the result to the broker, who verifies the proxy's identity and sends the response to the user.
\end{itemize}

\begin{table}
    \small
    \centering
    \caption{Query latency when using \system with ChatGPT.}
    \begin{tabular}{ccccc}
        \toprule 
        \thead{Activity} & \thead{Avg. time\\ (seconds)}  & \thead{Standard\\ deviation} & \thead{Margin\\ of error}  & \thead{Percent.\\ of total} \\
        \midrule
        \makecell{User delivering\\ query to proxy} & 3.77 & 1.79 & 0.67 & 24.47\% \\
        \makecell{Proxy interacting\\ with ChatGPT} & 8.30 & 1.24 & 0.46 & 53.86\% \\
        \makecell{Proxy delivering\\ response to proxy} & 3.34 & 1.98 & 0.74 & 21.67\% \\
        \midrule
        \makecell{Total time} & 15.41 & 3.58 & 1.34 & 100\% \\
        \bottomrule
    \end{tabular}
    \label{tab:perf_time}
\end{table}

\begin{figure}[ht!]
    \centering    \includegraphics[width=\linewidth]{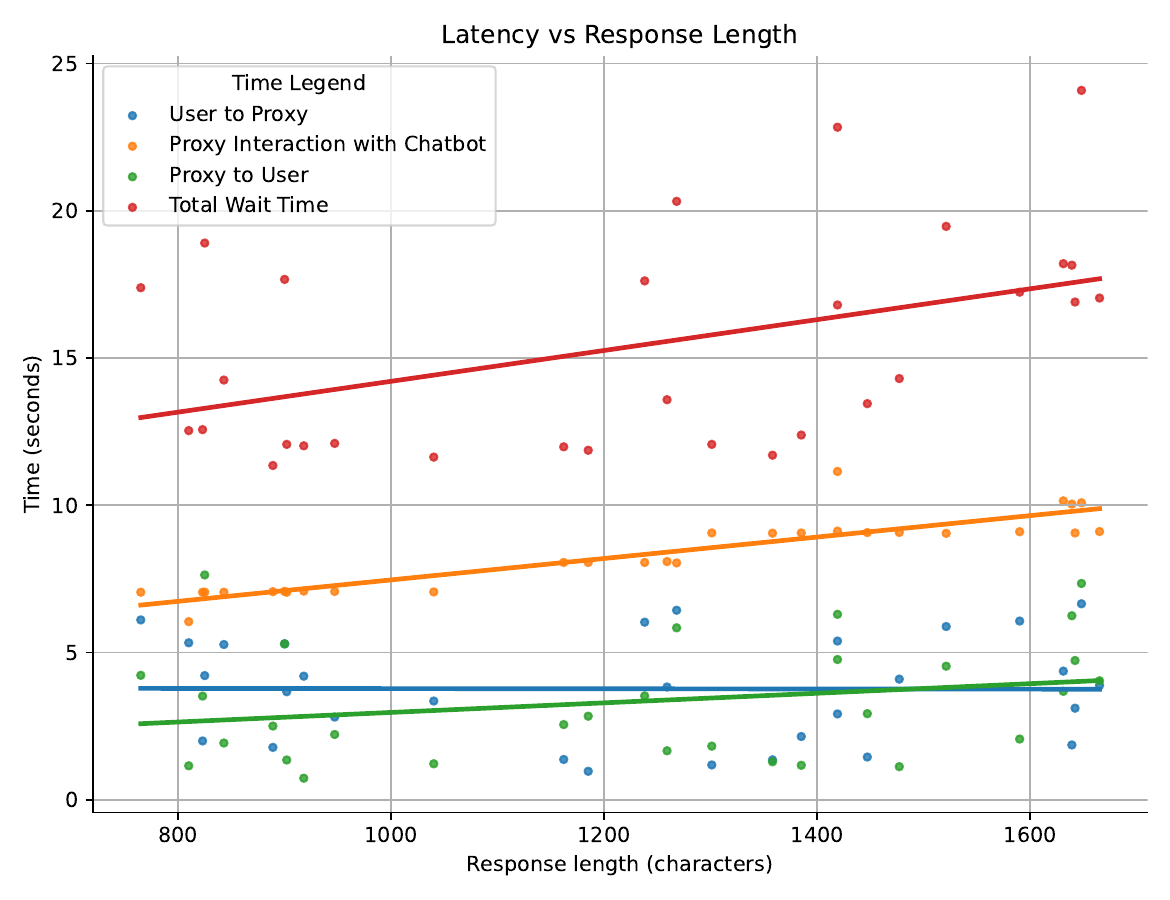}
    \caption{Breakdown of query latency vs ChatGPT's response length (in characters). Pearson's correlation coefficient between total wait time and response length is 0.4487, indicating a moderate positive correlation.}
    \label{fig:time_vs_response_length}
\end{figure}

Overall, it takes $15.41\pm1.34$ seconds on average for the user to receive the full response for a query, with roughly half of the delay ($8.3\pm0.46$ seconds) coming from the proxy's interaction with ChatGPT and the remaining half ($\approx 7.1$ seconds) coming from the Tor network communication between users, broker, and proxies (Table \ref{tab:perf_time}).
Of the $\approx8$ seconds spent interacting with ChatGPT's UI, about 1-2 seconds are idling time that we set to ensure that all UI operations and side-effects are fully propagated throughout ChatGPT's website internals correctly, while the remainder is from waiting for ChatGPT to return a complete response.
Thus, users only need to wait $<$10 additional seconds to get the full answer.
The longer the length of ChatGPT's response, the higher the latency (Figure \ref{fig:time_vs_response_length}), as the proxy has to spend more time waiting for the full response.
From these results, we estimate that a single proxy can serve 6-8 queries per minute, and a single user can ask 3-4 queries per minute.
This latency makes \system particularly suitable for more time-consuming chat features such as reasoning or Deep Research.

\subsection{Audit latency}
\begin{table}
    \small
    \centering
    \caption{TLSNotary latency (seconds) with ChatGPT.}
    \begin{tabular}{cccccc}
        \toprule 
        \thead{Distance from\\ notary (km)} & \thead{Mean}  & \thead{Median} & \thead{Max} & \thead{Standard\\ deviation} & \thead{Margin\\ of error} \\
        \midrule
        $\leq 1$ (no VPN) & 101.8 & 103 & 104 & 2.8 & 2.0 \\
        120 & 107 & 106 & 113 & 3.4 & 2.4 \\
        200 & 111.6 & 112 & 116 & 2.7 & 1.9 \\
        555 & 129.2 & 130 & 131 & 2.2 & 1.6 \\
        800 & 130.2 & 130 & 131 & 0.8 & 0.6 \\
        $>$800 & Timeout & N/A & N/A & N/A & N/A\\
        \bottomrule
    \end{tabular}
    \label{tab:audit_time}
\end{table}

We also run a small-scale measurement of the total amount of time for a proxy to finish integrity audits with TLSNotary and ChatGPT.
Note that proxies have to use a VPN instead of Tor to participate in the TLSNotary protocol due to Tor's high communication overhead.
To simulate the physical distance between a proxy and the notary server, we use Mullvad VPN~\cite{mullvad} and run 10 audits for each chosen VPN server.
On average, we find that the amount of time it takes to execute a single TLSNotary audit is typically between 100 and 130 seconds, with a moderate amount of variability depending on the proxy's network conditions (Table \ref{tab:audit_time}).
Longer distances between the proxy and the notary lead to longer latency and can cause the proxy to be unable to send its MPC-encrypted message to the chatbot server due to timeouts.
In an actual production deployment, the notary can be an independent service with multiple servers deployed globally to improve coverage and latency.
As TLSNotary is still under active development, we expect to see more optimizations for the protocol's reliability in the near future, which can potentially enable running it over the Tor network.

\subsection{In-browser LLM inference}

\subsubsection{Prompt safety}

We evaluate our browser-optimized Llama-Guard-3-1B model (Section \ref{sec:prompt_safety_impl}) on the AI safety dataset AEGIS 2.0~\cite{ghosh2025aegis}, specifically its test set, which contains 1928 samples (1039 unsafe, 889 safe) after filtering out invalid cases.
We compare the WebLLM version to an existing ONNX-converted q4f16 version for the Transformers.js engine.
(We only use q4f16 since the model fails to load for any larger sizes due to the browser's maximum memory allocation restriction.)
We test with the Google Chrome browser (version 136.0.7103.114) on two different hardware configurations: a high-end consumer desktop (Windows 11, 32GB RAM, i7-10700K CPU 3.8GHz, NVIDIA RTX 3090 GPU, Vulkan driver) and a low-mid range laptop (Windows 11, 16GB RAM, i7-1250U, Intel Iris Xe with 8GB shared memory).
We measure the prediction performance (precision, recall, F1 score), the inference speed per input token, and the storage/memory consumption.
From Table \ref{tab:safety_eval}, we can observe that the WebLLM is $\approx$30-50 times faster, two-thirds the size, and slightly better than Transformers.js.
The only downside is the more than doubled memory usage.
While the high-end dedicated GPU setup is $\approx$10-15 times faster than the integrated graphics, the latter's performance is still sufficient for real-time performance.

\begin{table}
    \setlength{\tabcolsep}{2pt}
    \small
    \centering
    \caption{Prediction performance, inference speed, and storage/memory usage of Llama-Guard-3-1B (q4f16) using WebGPU with different hardware and inference engines.}
    \begin{tabular}{cccccccccc}
        \toprule
        \multicolumn{2}{c}{Config} & \multicolumn{3}{c}{Prediction perf.} & \multicolumn{3}{c}{Speed (sec/tok)} & \multicolumn{2}{c}{Resources} \\
        \cmidrule(lr){1-2} \cmidrule(lr){3-5} \cmidrule(lr){6-8} \cmidrule(lr){9-10}
        Type & Engine & Prec & Rec & F1 & Avg & Med & Std & Size & RAM \\
        \midrule
        Desktop & WebLLM & 0.79 & 0.67 & 0.72 & 6e-04 & 6e-04 & 4e-05 & 0.8G & 6.2G  \\
        Desktop & Transformers & 0.70 & 0.69 & 0.70 & 0.03 & 0.03 & 0.02 & 1.2G & 2.9G \\
        Laptop & WebLLM & 0.78 & 0.66 & 0.72 & 0.01 & 0.01 & 3e-04 & 0.8G & 6.2G  \\
        Laptop & Transformers & 0.70 & 0.69 & 0.70 & 0.30 & 0.27 & 0.23 & 1.2G & 2.9G \\
        \bottomrule
    \end{tabular}
    \label{tab:safety_eval}
\end{table}

\begin{table}
    \setlength{\tabcolsep}{2pt}
    \small
    \centering
    \caption{Prediction performance, inference speed, and storage/memory usage of Piiranha-v1 using WebAssembly and Transformers.js with different hardware and quantizations.}
    \begin{tabular}{cccccccccc}
        \toprule
        \multicolumn{2}{c}{Config} & \multicolumn{3}{c}{Prediction perf.} & \multicolumn{3}{c}{Speed (sec/token)} & \multicolumn{2}{c}{Resources} \\
        \cmidrule(lr){1-2} \cmidrule(lr){3-5} \cmidrule(lr){6-8} \cmidrule(lr){9-10}
        Type & Quant. & Prec & Rec & F1 & Avg & Median & Std & Size & RAM \\
        \midrule
        Desktop & q4f16 & 0.958 & 0.834 & 0.894 & 0.014 & 0.012 & 0.005 & 0.4G & 1.2G \\
        Desktop & fp16 & 0.961 & 0.862 & 0.910 & 0.012 & 0.010 & 0.005 & 0.6G & 1.5G \\
        Desktop & fp32 & 0.961 & 0.862 & 0.910 & 0.010 & 0.007 & 0.004 & 1.2G & 2.9G \\
        Laptop & q4f16 & 0.958 & 0.834 & 0.894 & 0.011 & 0.009 & 0.004 & 0.4G & 1.3G \\
        Laptop & fp16 & 0.961 & 0.862 & 0.910 & 0.007 & 0.007 & 0.001 & 0.6G & 1.5G \\
        Laptop & fp32 & 0.961 & 0.862 & 0.910 & 0.006 & 0.005 & 0.001 & 1.2G & 2.9G \\
        \bottomrule
    \end{tabular}
    \label{tab:pii_eval}
\end{table}

\subsubsection{PII detection} \label{sec:eval_pii}

We evaluate the Piiranha-V1 model on the first 2000 validation samples of the Open-PII-Masking-500k dataset~\cite{ai4privacy_open_pii_masking_500k}.
For each quantization mode of the model, we measure its prediction performance (precision, recall, F1 score), inference speed, and resource usage (storage, CPU memory).
We test the same hardware configurations as in the prompt safety evaluation, but with the Tor browser (version 14.5.1) instead of Google Chrome since users interact with our website via Tor.
Results from Table \ref{tab:pii_eval} show that the model has good inference performance and speed for real-time usage, with the half-precision (fp16) model offering the best tradeoff between model size and prediction performance/speed.
Interestingly, the more quantized a model, the slower it becomes, which is likely because quantization on WebAssembly is not fully optimized.

\section{Usability Testing with Human Participants} \label{sec:human_eval}
To assess \system's usability, we run a semi‑structured user study, where users interact with the system using realistic queries. Participants highlight \system's main strengths, including suitability for privacy‑sensitive questions, an engaging proxy experience that motivates the use of our system, and an intuitive interface, but they also point out response latency as the main drawback.

\subsection{Setup}
We run a formative, semi‑structured study with five\footnote{Our testing sample of 5 participants aligns with the best practices established in previous literature.
\citet{virzi1992refining} shows that testing with 5--6 users typically uncovers 80-85\% of usability issues, with diminishing returns beyond this threshold.
\citet{10.1145/169059.169166} also shows that smaller, targeted samples achieve problem-discovery saturation more efficiently than larger single sessions. 
Since our system is under a small-scale research development, we determine a sample size of 5 to be reasonable as well as cost- and time-effective.} 
participants who alternate between two roles: (i) an end‑user submitting anonymous queries, and (ii) a volunteer proxy relaying and answering those queries. 
During the end‑user phase, every participant tests at least five queries based on three realistic but privacy‑sensitive scenarios. 
Immediately afterward, we interview each person for their assessment on the platform's usability and its relevance to their everyday usage. 
(See Appendix \ref{appendix:human-eval} for more details.)

\subsection{Findings}

\subsubsection{\system is highly suitable for use cases where users have sensitive queries} 
When participants use chatbots, their main concerns are that any PII they share might be stored indefinitely, repurposed for model training, or exposed in a breach. 
They are also concerned about query content privacy, as their conversation content could be combined with their personal information to build profiles for training or advertising purposes.
Although these platforms now offer ``delete conversation'' buttons and incognito modes, participants remain unconvinced that such measures truly remove all information stored on the server.

These anxieties shape how participants think about \system's values. 
All participants recommend \system in scenarios where users have sensitive queries, whether involving personally identifiable or sensitive information that they prefer not to be linked to their personal accounts.
Participants agree that \system's main advantage is that it does not require a user profile, which allows users to access all chatbot features while maintaining their privacy from chatbot providers. 

\subsubsection{Participants are mostly concerned about the latency of the platform, but consider this a trade-off for privacy}
All participants identify \system's response time as the main obstacle to regular use.
$P_2$ and $P_5$ consider latency mainly a user experience challenge that could be mitigated by streaming responses rather than returning them all at once.
Still, several interviewees view the delay as an acceptable tradeoff for stronger privacy protections: as $P_5$ put it, ``you spend more time waiting, but you can have the peace of mind that your queries are not traceable to your identity, which seems like a reasonable trade-off for me.''

\subsubsection{Proxy experience is interesting, though there is room for improvement}

With the exception of one participant, who never uses chatbots for sensitive queries and thus does not anticipate using \system, our participants enjoy the proxying experience and are willing to proxy to earn usage credit.
$P_3$ and $P_4$ appreciate the option to choose between multiple proxies, which can prevent any single proxy from viewing the entire conversation of a single user.
$P_5$ even describes this experience as ``fun,'' as they enjoy seeing the new queries that pop up on their chatbot screen.
None of the participants experience any interruption to their workflow or chatbot usage while serving as a proxy.
However, $P_2$, $P_3$, and $P_4$ express concerns about having random queries tied to their accounts, which could clutter their conversation history or cause legal issues, especially if the queries involve questionable activities.
The use of Llama Guard helps assuage their concerns somewhat, but not completely.

\subsubsection{\system interface is user-friendly}
All participants note that the interface is simple, clean, and intuitive.
$P_2$ and $P_5$ highlight the convenience of having both ChatGPT and Claude on the same platform, with $P_5$ mentioning that this will be particularly helpful for quickly experimenting with different chatbots to see which one gives a better answer.

\section{Privacy Analysis} \label{sec:analysis}

Here, we present an informal analysis of how well \system can defend against some de-anonymization attacks against its users.
Providing a fully formal and provable guarantee of \system's privacy protection is infeasible without making unrealistically idealized assumptions.
Even well-established AC systems like Tor do not offer provable end-to-end unlinkability~\cite{karunanayake2021torsurvey}.
We assume the same adversary \adversary as described in our threat model (Section \ref{sec:threat_model}).

\subsection{Traffic Analysis}

\adversary as it is can only directly observe the traffic from its controlled chatbot \chatbot's regular users and known \system proxies.
\system users always use an AC protocol to communicate with other parties, except when they need to volunteer as proxies to earn our e-cash.
This is the only occasion where \adversary can observe the users' traffic.
Thus, one possible linkage attack that \adversary can perform is as follows:
\begin{itemize}
    \item \adversary pretends to be a \system user and sends queries meant for \chatbot to a proxy \proxy. This continues until \adversary detects that the \proxy participates in an integrity audit with \chatbot.
    \item \adversary collects all user queries sent from other proxies after the audit finishes.
    \item \adversary matches the collected queries with the profile of \proxy based on the queries' timing or content.
\end{itemize}

This attack can work if \proxy works with the controlled \chatbot and it uses its earned e-cash (from proxying for \adversary) to also ask a query for \chatbot.
However, with e-cash isolation (i.e., e-cash earned from proxying for one chatbot cannot be spent for the same chatbot), this linkage is no longer feasible since \adversary cannot observe traffic belonging to a different chatbot platform.
We believe the assumption that \adversary can only control a single chatbot is realistic because of the competition between different chatbot companies.

\subsection{Content Analysis}

By \system's nature, the system does not prevent the chatbot provider from de-anonymizing users using the content of their queries (Section \ref{sec:overview}).
However, it can readily incorporate ML-based techniques to assist users with sanitizing their queries (Section \ref{sec:content_privacy}), not to mention the ease with which it can switch to newer and better ML models as they arrive.
Moreover, content analysis is unlikely to work on a user if they do not already have an account with the chatbot controlled by \adversary (unless \adversary also collects external data beyond the chatbot conversations).
Our e-cash isolation scheme further reinforces this separation.
Even if \adversary can determine that a group of queries belongs to a user, as long as the user's identity remains hidden, the cross-query linkage is also unlikely to help with de-identification.

\subsection{Chatbot Response Manipulation}

\adversary can maliciously modify the chatbot responses to expose users' identities.
For example, \adversary may inject a unique watermark~\cite{kirchenbauer2023watermark} into the response for each query coming from known \system proxies.
If a user accidentally includes the watermark in their non-\system-protected chats, then \adversary can link the user's identity to the watermarked query.
A more direct attack is by including malicious code or links that, when run/clicked, would expose the user's identity, similar to cross-site scripting attacks.
This approach, however, is too easily detectable and can harm the chatbot's reputation.
Overall, defending against malicious response manipulation is not straightforward, even with systems like SMC or TEE, since we cannot control how users make use of the chatbot responses.
We believe a promising approach is for users to use LLMs to process the responses, e.g., by rewriting to remove any watermark and checking for malicious content.

\section{Discussion}

Here, we examine \system's strengths \& limitations, ethical concerns, future extensions, and related work.

\subsection{Strengths and Limitations} \label{sec:limitations}

Table \ref{tab:comparison} highlights several key differences between \system and existing solutions to private LLMs, particularly SMC, TEE, and VPN-like chatbots such as DuckAI and AnonChatGPT.
Overall, \system has comparatively good speed, provides user identity separation, requires no modifications to the chatbots, supports different chatbots, and also verifies request/response integrity.

\begin{table}[h]
    \setlength{\tabcolsep}{2pt}
    \footnotesize
    \centering
    \caption{Pros and cons of different approaches to private LLM}
    \begin{tabular}{ccccccc}
        \toprule
        \thead{Method} &  \thead{Service\\ latency} & \thead{Identity\\ separation?} & \thead{Chatbot\\ modification?} & \thead{Chatbot\\ restriction?} & \thead{Content\\ integrity?} & \thead{Content\\ privacy?}\\
        \midrule
        \makecell{SMC\\ \cite{pang2024bolt, lu2025bumblebee, key2025shaft}} & \cellcolor{Dandelion}Slow & \cellcolor{Dandelion}No & \cellcolor{Dandelion}Yes & \cellcolor{Dandelion}Yes & \cellcolor{LimeGreen}Yes & \cellcolor{LimeGreen}Partial \\
        \makecell{TEE\\ \cite{zhu2024tee, applepcc, meta_private_processing}} & \cellcolor{LimeGreen}Fast & \cellcolor{Dandelion}No & \cellcolor{Dandelion}Yes & \cellcolor{Dandelion}Yes & \cellcolor{LimeGreen}Yes & \cellcolor{LimeGreen}Partial \\
        \makecell{DuckAI, etc.\\ \cite{duckai, anonchatgpt, leoai}} & \cellcolor{LimeGreen}Fastest & \cellcolor{LimeGreen}Partial & \cellcolor{LimeGreen}No & \cellcolor{Dandelion}Yes & \cellcolor{Dandelion}No & \cellcolor{Dandelion}No \\
        \makecell{\system\\ (\textbf{ours})} & \cellcolor{LimeGreen}Fast & \cellcolor{LimeGreen}Yes & \cellcolor{LimeGreen}No & \cellcolor{LimeGreen}No & \cellcolor{LimeGreen}Yes & \cellcolor{Dandelion}No \\
        \bottomrule
    \end{tabular}
    \label{tab:comparison}
\end{table}

\subsubsection{Latency}
\system's increased latency of 7-8 seconds over normal conversations is mainly dominated by the delay from the AC protocol (and the chatbot service itself).
While faster alternatives to Tor based on OHTTP~\cite{rfc9458ohttp} or MASQUE~\cite{masque, dikshit2023masque} can potentially be used, they are still being developed and are not as secure as Tor~\cite{zohaib2023icloud}.
\system's speed can also be influenced by the proxy integrity audit, whose latency depends on the underlying web proof protocol.
There is active research on faster web proofs~\cite{celi2025distefano, ernstberger2024origo, luo2024proxyingisenough}, but few publicly available frameworks.
Regular conversations will see a larger impact from \system's delay, but time-consuming features such as Deep Research~\cite{chatgpt_deepresearch}, which can take 5-10 minutes to finish a single query, are particularly suitable for the system.

\subsubsection{User identity}

\system's privacy protection is only as good as the components used in the system.
In particular, Tor is known to be vulnerable to adversaries capable of global traffic analysis~\cite{karunanayake2021torsurvey},
TLSNotary is an emerging protocol that has not been battle-hardened, and Chaumian e-cash relies on the hardness of the RSA problem~\cite{bellare2003blindsig}.
Nonetheless, our identity separation is still more thorough than chatbots like DuckAI, which operate similarly to a VPN.
Content-based linkage attacks might be feasible in theory, but without our decoupling of the queries and their owners, user identities would always be visible to the providers.

\subsubsection{Dependence on chatbot providers}
\system is completely independent of any chatbot providers, unlike SMC and TEE, which require special modifications to chatbots.
This independence allows the system to flexibly extend its support to a variety of different, competing chatbots.
As a trade-off, we do not have any guarantees for the integrity of the providers, who may punitively alter or restrict their services for suspected \system participants.

\subsubsection{Content integrity} \label{sec:proxy_integrity_analysis}

Our proof-of-concept probabilistic proxy integrity audits can prevent proxies from misbehaving to an extent, but it does not completely rule out such foul play.
A malicious proxy could try to distinguish audit queries from real user queries and only act honestly when handling the former, but this can be practically prevented with careful query generation (Section \ref{sec:audit}).
Assuming audit queries are indistinguishable from regular ones, we can model this scenario as a 2-player non-cooperative game~\cite{nash1951game} (Appendix \ref{apd:proxy_integrity}).
As web proof technology becomes faster, we can switch from probabilistic to mandatory audits to better ensure integrity.

\subsubsection{Content privacy}

\system does not prevent the chatbot providers or the proxies from accessing the plaintext of users' conversations.
To remedy this, the system could be combined with a TEE-based solution to achieve both strong cryptographic content and practical identity hiding.
For example, Apple's Private Cloud Compute~\cite{applepcc} proposes using a third-party OHTTP~\cite{rfc9458ohttp} relay to send users' chat requests to its TEE.
That being said, this requires extensive support from the chatbot providers, the majority of which likely do not prioritize user privacy enough to include this extra complexity.
Moreover, TEE or SMC for LLMs cannot be easily extended to features like LLM web search, which involves querying the internet with information from the user queries.

\subsection{Ethical Considerations} \label{sec:ethics}

\subsubsection{Terms of service}

Most chatbot providers currently prohibit automated access to their services except through officially supported APIs.
Consequently, our browser extension’s method of site interaction may be interpreted as a policy violation.
However, these providers typically grant users full ownership of the content generated during their interactions, and our system only accesses data with the users’ explicit consent.
This restriction highlights a broader imbalance: while chatbot providers routinely collect large-scale web data without obtaining individual consent, they seek to limit users’ ability to control or protect their own data.
We maintain that empowering users with privacy-preserving tools is at least as ethically defensible as the extensive web scraping employed by chatbot providers to train their own LLMs~\cite{nyt2023lawsuit}.
Furthermore, our implementation is strictly a research-oriented proof of concept, not intended for commercial deployment.
Additionally, our give-and-take economy helps reduce strain on chatbot services by discouraging excessive automated usage.

\subsubsection{Misuse}
As with any anonymity services, \system can be exploited by malicious actors for harmful purposes~\cite{openai2024malicious}, potentially implicating the proxies.
Although prompt safety models like Llama-Guard-3-1B are not foolproof~\cite{wei2024jailbroken}, equipping proxies with such defenses allows them to selectively assist appropriate content.
We argue that the primary responsibility for aligning LLMs with societal values lies with chatbot providers, who must prevent harmful responses at the model level~\cite{ouyang2022align}.
Addressing misuse through safety alignment is also likely more effective than attempting to police human online activities.
Ultimately, although misuse must be managed through thoughtful safeguards, the right to anonymity should not be sacrificed, as it plays a vital role in protecting individual dignity and enabling free, democratic discourse.


\subsection{Future Extensions}

\subsubsection{More decentralization} \label{sec:decentralization}
Our proof of concept implements the communication broker as a Tor hidden server for demonstration purposes.
However, the system's design does not exclude a multi-broker architecture to achieve more extensive decentralization, e.g., similar to Tor's Directory Authorities.
Completely decentralizing every trusted component, however, is challenging due to the proxy integrity requirements of our system.
For instance, although replacing the current trusted notary approach with peer-to-peer notarization via TLSNotary is possible, this requires careful execution to ensure the underlying WebRTC-based peer-to-peer communication does not leak the IP of the involved parties.
Running TLSNotary over Tor is not currently practical due to the high network overhead caused by the interactive nature of the protocol.
Even if peer-to-peer web proof is supported, we still need to rely on trusted audits for our give-and-take economy to prevent proxies from gaming the reward system via collusion with users.
That said, we believe a fully decentralized \system is worth pursuing to further promote users' trust and improve the system's resilience.

\subsubsection{Additional chatbot features}
We intend to incorporate more chatbots, such as Gemini and Meta AI, to attract even more prospective users.
Furthermore, we hope to extend beyond purely natural text conversations to multi-modal chat features (e.g., files, images, and audio).
One obstacle to including more functionalities is the manual coding involved to customize for different and often changing chatbot UIs.
As such, we also plan to investigate how to utilize LLMs to automatically explore a chatbot UI and dynamically add new features for both proxies and users.

\subsection{Related Work} \label{sec:background}

\subsubsection{Content-based privacy techniques.} \label{sec:content_privacy}

Recent research attempts \emph{prompt anonymization} using LLMs to analyze and sanitize user prompts for privacy leakage beyond PII~\cite{chong2024casper, siyan2025papillon, zhou2025rescriber}.
LLMs are also applied to perform \emph{authorship obfuscation} to hide the linkage between a text's characteristics and its author~\cite{fisher2024jamdec, fisher2024styleremix, xing2024alison}.
While these methods can be incorporated into \system to help improve the privacy of user prompts' content, there are several limitations: First, good performance requires sufficiently powerful LLMs, severely constraining users' computing resources.
Second, ML-based techniques do not provide any formal guarantee and can exhibit unexpected failure modes~\cite{pham2025names}.
Nonetheless, this approach has great potential as LLMs become increasingly better at smaller sizes.

\subsubsection{Proxy-based anonymity.}
One of the most notable proxy-based systems is The Onion Router project (Tor), featuring a decentralized network of volunteer relays that perform onion routing~\cite{dingledine2004tor}.
Other proxy-based censorship circumvention applications include Snowflake~\cite{bocovich2024snowflake}, MassBrowser~\cite{nasr2020massbrowser}, and Hola VPN~\cite{hola}, which all make use of peer browser volunteers to proxy traffic.
While similar in spirit, \system does not proxy arbitrary internet traffic, but rather high-level interactions with credential-required chatbots.
We borrow the term `interaction proxy' from the human-computer interaction community, which uses it to refer to a layer that interacts with a physical device on behalf of users~\cite{huang2023interaction, lu2024interactout}.
The idea of using proxying for high-level web services can also be traced back to the early days of the web with systems like Crowds~\cite{reiter1998crowds}, Janus~\cite{gabber1997janus}, LPWA~\cite{gabber1999lpwa}, and UPIR~\cite{domingo2009upir}.
However, they assume non-malicious peers, which is not realistic for our use case.

\section{Conclusion}

In this paper, we present \system, a novel proxy-based privacy-enhancing system for LLM chatbot users that leverages the power of volunteers to separate a user's identity from their queries.
Our system utilizes TLSNotary for proxy integrity, Llama Guard for prompt safety, and Chaum's e-cash for sustainability.
We envision \system as a platform that can augment the capability and popularity of chatbots rather than circumventing them, thus bringing their services to an even broader audience who might feel hesitant due to privacy concerns. 
We hope that our work will inspire chatbot providers to build their services with privacy at the forefront, especially as this powerful technology becomes more intertwined with everyday lives.


\newpage

\section*{Acknowledgment}
The authors used generative AI-based tools to revise some portions of text, mainly to improve flow and correct typos, grammatical errors, and awkward phrasing.

\bibliographystyle{ACM-Reference-Format}
\bibliography{references}

\appendix

\section{Proxy Integrity Analysis} \label{apd:proxy_integrity}

Consider the following reward matrix for a malicious proxy and a broker:
\begin{table}[h!]
    \small
    \centering
    \caption{Generic Reward Matrix. The numbers on the left of the tuples are the rewards for the broker, and the ones on the right are for the proxy.}
    \begin{tabular}{|c|c|c|}
        \hline
        \diagbox[height=3\line]{Coord.}{\\Proxy} & Honest & Dishonest \\
        \hline
        \multirow{2}{*}{Audit} & \multirow{2}{*}{$(r^{c}_{ah}, r^{p}_{ah})$} & \multirow{2}{*}{$(r^{c}_{ad}, r^{p}_{ad})$} \\
        & & \\
        \hline
        \multirow{2}{*}{Non-audit} &  \multirow{2}{*}{$(r^{c}_{nh}, r^{p}_{nh})$} & \multirow{2}{*}{$(r^{c}_{nd}, r^{p}_{nd})$} \\
        & & \\
        \hline
    \end{tabular}
    \label{tab:reward_matrix_2}
\end{table}

We can simplify the matrix by fixing the rewards of the worst and best possible outcomes for each player.
More specifically, the malicious proxy wins if it acts dishonestly with a non-audit query ($r^{p}_{nd} = 1)$ and loses if the query is an audit one ($r^{p}_{ad} = -1$).
The broker wins if it catches the proxy being dishonest ($r^{c}_{ad} = 1$) and loses if it fails to ($r^{c}_{nd} = -1$).
In addition, the broker does not get any reward if it does not perform an audit and the proxy is honest ($r^{c}_{nh} = 0$).

\begin{table}[h!]
    \small
    \centering
    \caption{Simplified Reward Matrix.}
    \begin{tabular}{|c|c|c|}
        \hline
        \diagbox[height=3\line]{Coord.}{\\Proxy} & Honest & Dishonest \\
        \hline
        \multirow{2}{*}{Audit} & \multirow{2}{*}{$(r^{c}_{ah}, r^{p}_{ah})$} & \multirow{2}{*}{$(1, -1)$} \\
        & & \\
        \hline
        \multirow{2}{*}{Non-audit} &  \multirow{2}{*}{$(0, r^{p}_{nh})$} & \multirow{2}{*}{$(-1, 1)$} \\
        & & \\
        \hline
    \end{tabular}
    \label{tab:reward_matrix_3}
\end{table}

We further assume the following inequalities:
\begin{itemize}
    \item $-1 < r^{c}_{ah} \leq 0$: Performing audits can cost the broker, but is less costly than missing a dishonest query.
    \item $r^{p}_{ah} \geq 0$: Being honest with an audit query can reward the proxy even though the proxy has to spend extra computation resources to complete the audit.
    \item $-1 < r^{p}_{nh} \leq 0$: Being honest with a non-audit query can be costly since the proxy has to spend computation resources without being rewarded.
    However, it is less costly than being caught dishonest and banned from participating.
\end{itemize}

Let $p_a$ be the probability of auditing, and let $p_h$ be the probability of the proxy being honest.
The state where no players can improve their utility by changing their strategy is called \emph{Nash Equilibrium} (NE).
A pure-strategy NE is when the actions are chosen deterministically, whereas a mixed-strategy NE is when the actions are chosen stochastically.

We can see that no pure-strategy NE exists with the simplified reward scheme.
The proxy's expected reward is $r^p_{ah} p_a + r^p_{nh} (1 - p_a)$ if it is honest and $-p_a + (1 - p_a) = 1 - 2p_a$ if it is dishonest.
Thus, the proxy will mix between the two strategies if:
$$r^p_{ah} p_a  + r^p_{nh} (1 - p_a) = 1 - 2p_a = \iff p_a = \frac{1 - r^p_{nh}}{r^p_{ah} - r^p_{nh} + 2}$$

The broker's expected reward is $r^c_{ah} p_h + (1 - p_h)$ if it audits and $p_h - 1$ if it does not.
Thus, the broker will mix between the two strategies if:
$$r^c_{ah} p_h + (1 - p_h) = p_h - 1 \iff p_h = 2/(2 - r^c_{ah})$$

Therefore, the mixed-strategy NE is $p^*_a = 1/(r^p_{ah} - r^p_{nh} + 2)$ and $p^*_h = 2/(2 - r^c_{ah})$. Since we assume that $-1 < r^c_{ah} \leq 0$, we have $2/3 < p^*_h \leq 1$.

The mixed-strategy NE expected reward for the proxy is:
$$E[R_P] = p^*_a (p^*_h r^p_{ah} - (1 - p^*_h)) + (1 - p^*_a) (p^*_h r^p_{nh} + (1 - p^*_h))$$

The mixed-strategy NE expected reward for the broker is:
\[
E[R_C] = p^*_a (p^*_h r^c_{ah} + (1 - p^*_h)) - (1 - p^*_a) (1 -p^*_h) = \frac{r^c_{ah}}{2 - r^c_{ah}}
\]
Since $-1 < r^c_{ah} \leq 0$, we have $-1/3 < E[R_C] \leq 0$.

Depending on the exact rewards, we have the following results for the mixed-strategy NE and expected rewards:
\begin{table}[h!]
    \centering
    \tiny
    \setlength{\tabcolsep}{2pt}
    \caption{Mixed-strategy NE and expected payoff for different reward schemes}
    \begin{tabular}{ccccc}
        \toprule
        Scenario & $p^*_a$ & $p^*_h$ & $E[R_P]$ & $E[R_C]$ \\
        \midrule
        $r^{p}_{ah} = r^{p}_{nh} = r^c_{ah} = 0$ & $1/2$ & $1$ & $0$ & $0$\\
        \midrule
        $r^{p}_{ah} = r^{p}_{nh} = 0, r^c_{ah} < 0$ & $1/2$ & $\displaystyle\frac{2}{2 - r^c_{ah}}$ & $0$ & $\displaystyle \frac{r^c_{ah}}{2 - r^c_{ah}}$ \\
        \midrule
        $r^{p}_{ah} > 0, r^{p}_{nh} = 0$ & $\displaystyle \frac{1}{r^{p}_{ah} + 2}$ & $\displaystyle\frac{2}{2 - r^c_{ah}}$ & $\displaystyle \frac{r^{p}_{ah}}{r^{p}_{ah} + 2}$ & $\displaystyle \frac{r^c_{ah}}{2 - r^c_{ah}}$ \\
        \midrule
        $r^{p}_{ah} = 0, r^{p}_{nh} < 0$ & $\displaystyle \frac{1 - r^{p}_{nh}}{2 - r^{p}_{nh}}$ & $\displaystyle\frac{2}{2 - r^c_{ah}}$ & $\displaystyle \frac{r^{p}_{nh}}{2 - r^{p}_{nh}}$ & $\displaystyle \frac{r^c_{ah}}{2 - r^c_{ah}}$ \\
        \midrule
        $r^{p}_{ah} > 0, r^{p}_{nh} < 0$ & $\displaystyle \frac{1 - r^p_{nh}}{r^p_{ah} - r^p_{nh} + 2}$ & $\displaystyle\frac{2}{2 - r^c_{ah}}$ & $\displaystyle \frac{r^{p}_{ah} + r^{p}_{nh}}{r^p_{ah} - r^p_{nh} + 2}$ & $\displaystyle \frac{r^c_{ah}}{2 - r^c_{ah}}$ \\
        \bottomrule
    \end{tabular}
    \label{tab:mixed_ne}
\end{table}

Based on this, we can draw the following high-level guidelines when designing and implementing \system:
\begin{itemize}
    \item To reduce the need for audits (i.e., decrease $p^*_a$), we can increase the reward for proxies for successfully completing audits $r^p_{ah}$, such as by increasing the number of e-cash.
    Decreasing the cost of honest proxying $r^p_{nh}$ is less straightforward since this is often a fixed constant (e.g., each proxied request costs the proxy one chatbot request).
    With careful design, we could focus on the unspent request bandwidth that would have been wasted.
    \item To encourage proxies to be honest (i.e., increase $p^*_h$), we can decrease $r^c_{ah}$, the cost for the broker to perform audits when the proxy is honest.
    This is largely dependent on the underlying TLS-backed data provenance protocol.
\end{itemize}

We note that the current analysis model only allows us to make relative interpretations concerning the behavior of the proxies and the broker.
It does not enable us to determine the exact amount of the rewards needed to achieve a beneficial scenario.
We would need a more detailed model that can, for example, correctly relate the two quantities $r^p_{ah}$ and $r^c_{ah}$ since giving more rewards to a proxy can introduce future costs to the system.
We leave this for future work.


\section{Additional Discussion}

\subsection{Strawman Solutions} \label{sec:strawman}

During the development of \system and the writing of this paper, we received several suggestions on seemingly simpler alternatives to our proxy-based approach.
Here, we aim to elaborate on these strawman ideas and clarify their weaknesses in comparison to \system.

\subsubsection{Burner account}
One idea is for users to create temporary chatbot accounts using throwaway credentials.
Many online services provide temporary email addresses and phone numbers for this purpose.
However, chatbot services can easily block such credentials (e.g., by checking the email domain names and the phone number carriers), thus requiring users to manually create or obtain more long-term ones to set up their burner accounts.
This process does not scale well if users need anonymity frequently.

\subsubsection{Account sharing}
Another idea is for multiple users to share the same chatbot account.
There are a few websites that offer this type of service, such as TeroBox\footnote{\url{https://terobox.com}} and OPKFC\footnote{\url{https://www.opkfc.com}}, where users can discover or contribute their accounts.
From what we can gather, there are two types of shared accounts: public (i.e., credentials are available) and hidden (i.e., functionalities are abstracted behind a UI, often a mirror site).
Publicly shared accounts (e.g., TeroBox) can be easily blocked by chatbot providers because there are too many locations from which users can log in.
Hidden shared accounts (e.g., OPKFC), similar to chatbots like DuckAI, do not guarantee integrity and often rely on API usage.

\begin{figure}[h]
    \centering
    \begin{subfigure}[b]{0.48\linewidth}
        \includegraphics[width=\linewidth]{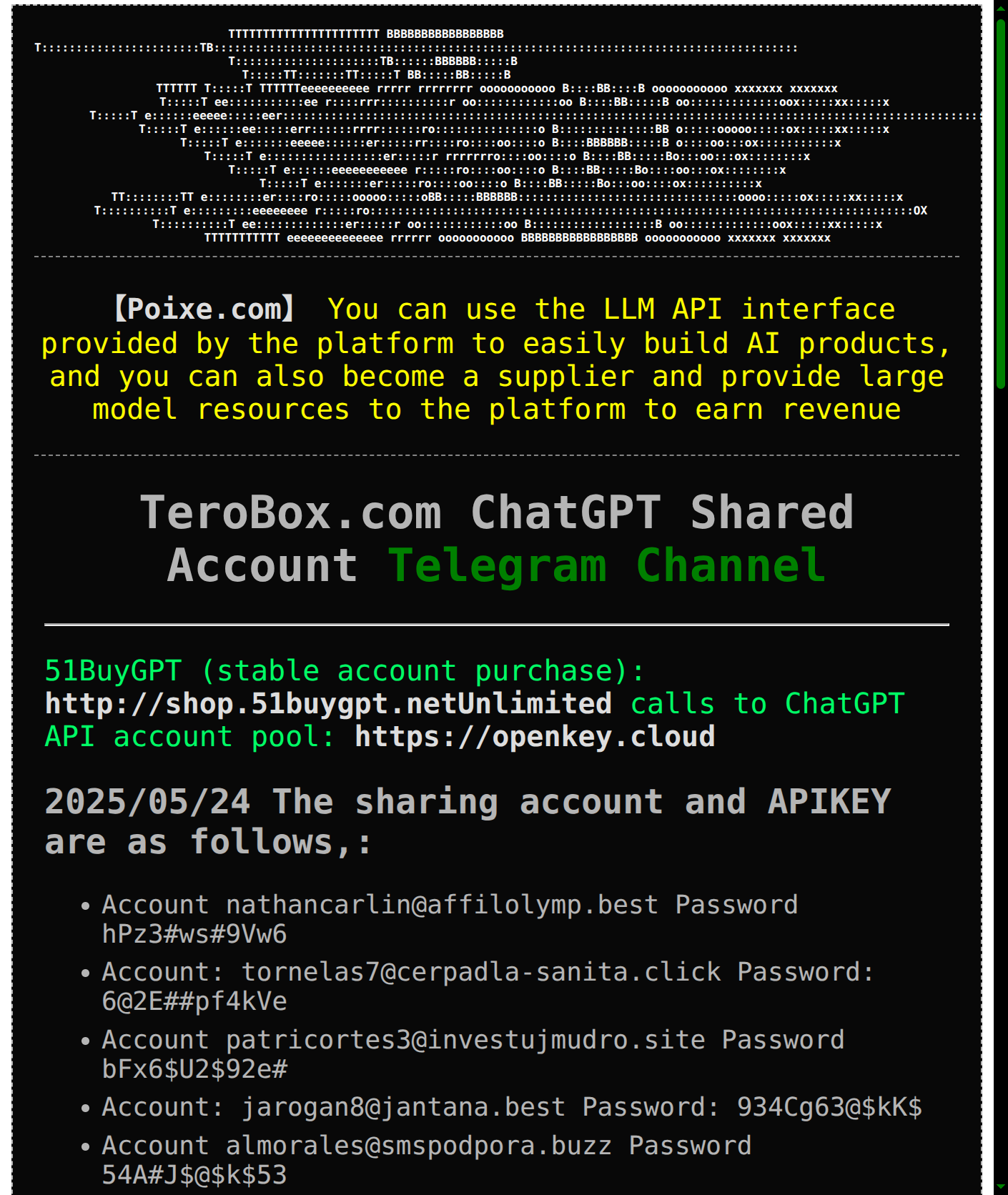}
        \caption{TeroBox}
        \label{fig:terobox}
    \end{subfigure}
    \hfill
    \begin{subfigure}[b]{0.48\linewidth}
        \includegraphics[width=\linewidth]{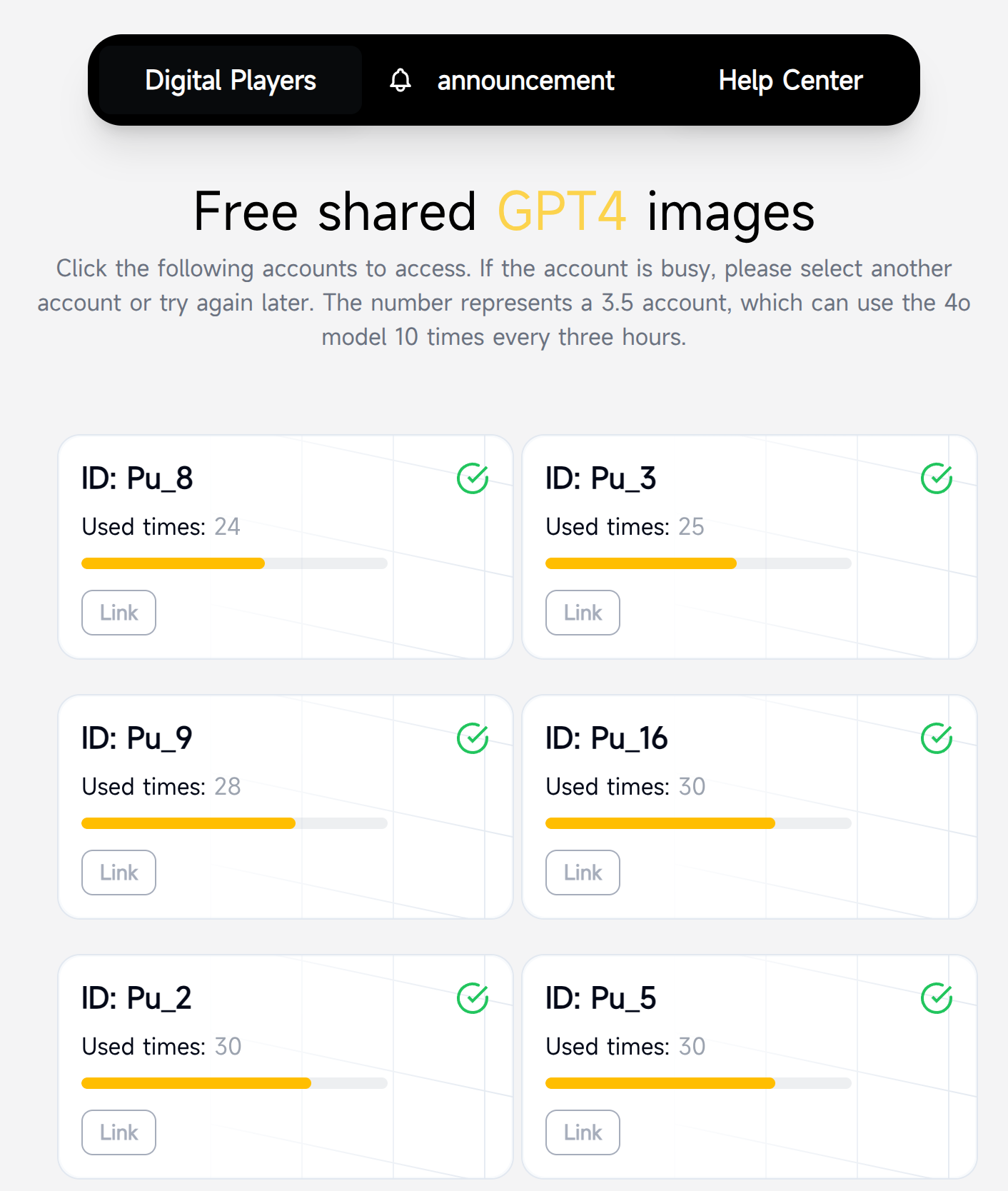}
        \caption{OPKFC}
        \label{fig:opkfc}
    \end{subfigure}    
    \caption{Examples of ChatGPT account sharing sites (auto-translated by Google Translate from Chinese to English).}
    \label{fig:account_sharing}
\end{figure}

\subsubsection{API proxy}
These proxies use the chatbots' API instead of directly interacting with the web UI.
While faster, this approach costs real money to run.
If any service offers this feature for free, then more often than not, the users are the product.
For instance, AnonChatGPT~\cite{anonchatgpt} serves ads on their website (not to mention the embedded Google Analytics script, which is certainly not privacy-friendly).

\subsubsection{AC protocol with loginless mode}
For chatbots that do not always require logging in, users could hypothetically use an AC protocol like Tor or a VPN to access those services anonymously.
In reality, this is often disallowed by the chatbots, as blocking Tor's exit nodes or suspected VPN IPs is fairly straightforward.
For example, both ChatGPT and Gemini disable the loginless feature when we use Tor to access them.

\begin{figure}[h]
    \centering
    \begin{subfigure}[b]{0.48\linewidth}
        \includegraphics[width=\linewidth]{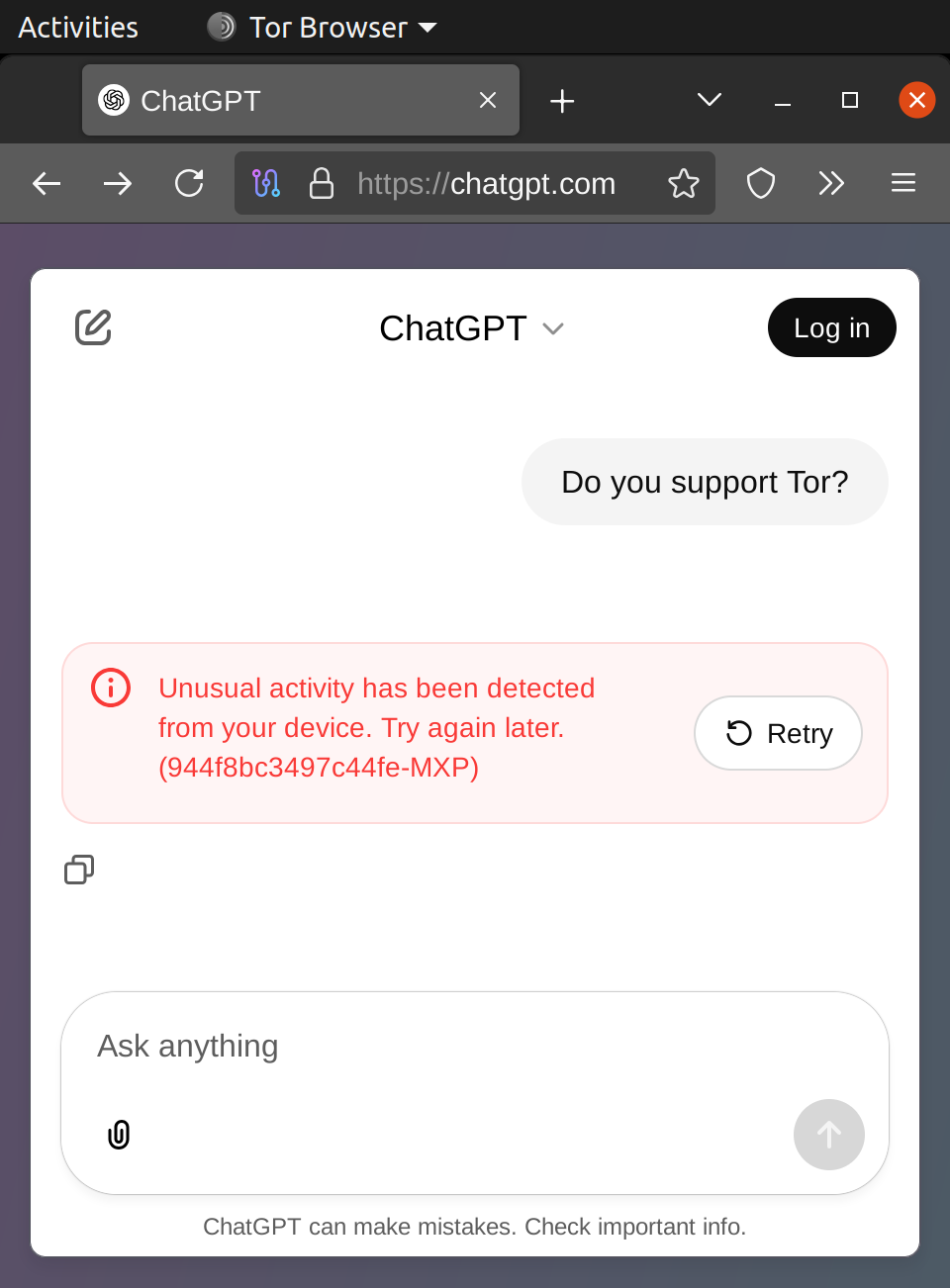}
        \caption{ChatGPT}
        \label{fig:chatgpt_tor}
    \end{subfigure}
    \hfill
    \begin{subfigure}[b]{0.48\linewidth}
        \includegraphics[width=\linewidth]{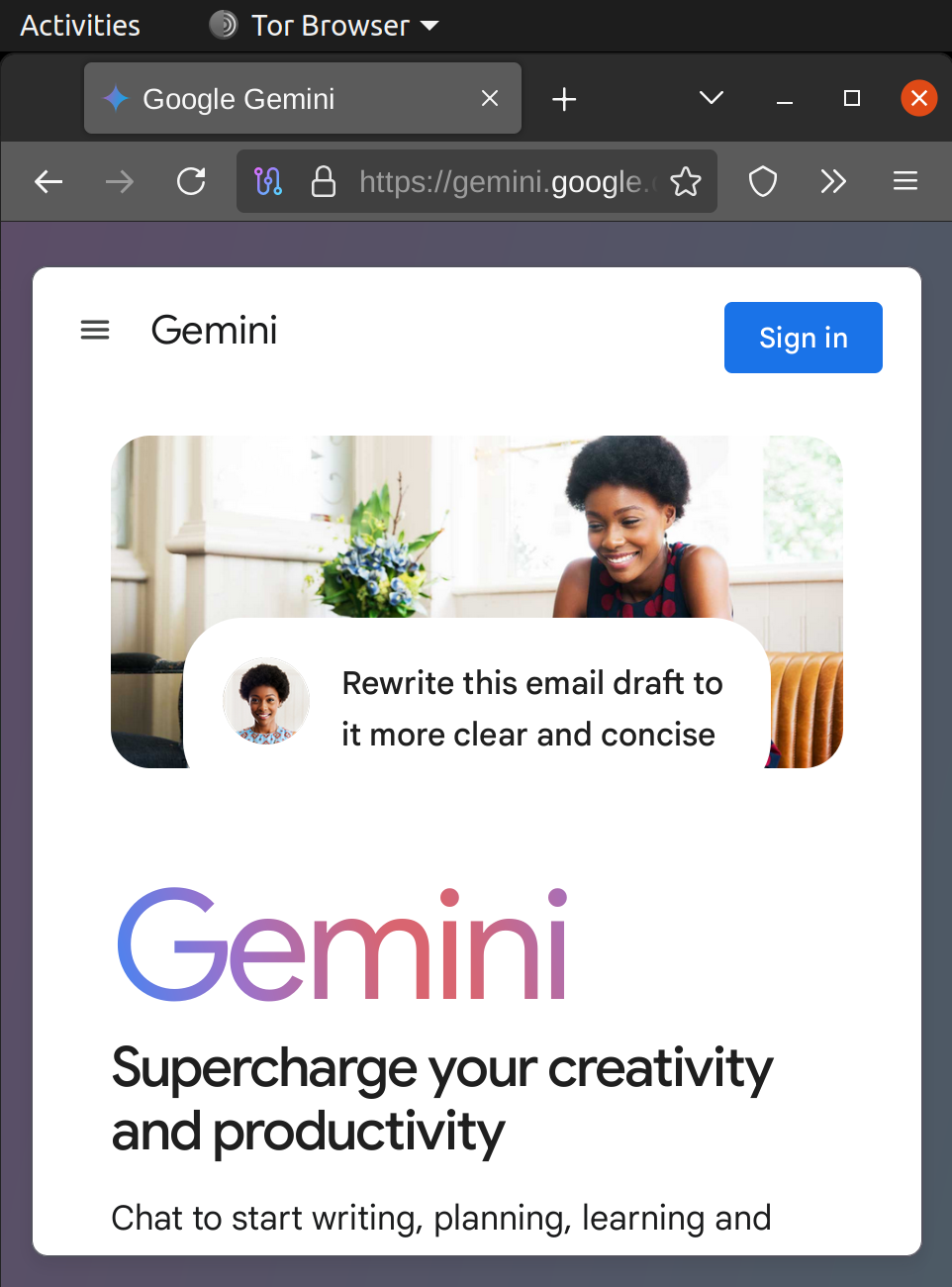}
        \caption{Gemini}
        \label{fig:gemini_tor}
    \end{subfigure}    
    \caption{ChatGPT and Gemini do not currently enable loginless mode for Tor.}
    \label{fig:chatbot_tor}
\end{figure}

\subsection{Human Evaluation}
\label{appendix:human-eval}
The five participants come from diverse backgrounds: two researchers specializing in LLMs ($P_1$, $P_2$), two researchers in security and privacy ($P_3$, $P_4$), and one layperson without research expertise ($P_5$). All participants use ChatGPT or Claude regularly (ranging from weekly to daily) for tasks such as coding, writing assistance, or general knowledge inquiries.

To minimize potential bias, we select individuals with no prior knowledge of \system. We enroll people from the same organization and professional network, and avoid recruiting people from crowdsourcing platforms due to logistical challenges.\footnote{We require participants to install Tor and our Chrome extension on personal laptops. Since ProxyGPT is an early prototype, this process demands troubleshooting and security clearances from our institution.}

The one-hour interviews are conducted by two interviewers, either in person or via Zoom for remote participants. Prior to the interview, we obtain consent from participants via email. Interview transcripts are manually anonymized, and recordings are set to be destroyed within one year after the interview. Upon completion, each participant receives a \$20 Amazon gift card. Our study is approved by our institution's institutional review board (IRB).

Here is the list of scenarios in our interviews:
\begin{itemize}
\item You began your job without obtaining proper legal authorization, and now you might face deportation. You plan to compile a preliminary list of resources before consulting costly lawyers.
\item You suspect that you might have a highly stigmatized disease (e.g. a sexually transmitted disease). You plan to research symptoms, explore treatment options, and seek empathy from external sources.
\item You need advice on managing student loans. You plan to get an overview of debt relief options, learn about budgeting techniques, and find supporting resources.
\end{itemize}
To analyze the interview data, we use a bottom-up thematic analysis approach~\cite{strauss1994grounded}, where the interviewers review recorded notes and sort participant responses into major themes.

\subsection{Practical Security Concerns}

\subsubsection{DoS and Sybil attacks}

In addition to the built-in DoS prevention mechanisms in the Tor network~\cite{tordos}, our give-and-take e-cash economy can also be considered a form of DoS defense thanks to its reliance on completing TLSNotary-based audits, which have high computational costs.
Sybil attacks, however, are harder to prevent due to the anonymous nature of \system.
If a dedicated actor floods our system with a large number of proxies, they can collect a large amount of e-cash over an extended period of time, which can then be used to ask many queries.
Consequently, they can collect a sizeable volume of user queries as well as chatbot responses and also affect the reputation and load of other proxies.
The use of e-cash and TLSNotary can prolong the time it takes for an adversary to achieve such an influence on our system.

\subsubsection{Key storage in browsers}

Securely storing sensitive information on the browser side is generally difficult to achieve completely.
Our browser extension relies on current native browser technologies, particularly the Web Crypto API and the IndexedDB API, to generate and store private keys locally on the browser side~\cite{pomcor2017keys}.
While this approach prevents the private keys from being exported to more accessible formats in JavaScript, it does not prevent malicious actors with access to the proxy's browser from using the key or extracting information from the raw binary using other tools.
If a cross-site scripting (XSS) attack is successfully executed on the extension, then the private keys can be stolen and abused.
To reduce the risk of XSS, our implementation sanitizes any inputs coming from users and proxies before displaying them with DOMPurify~\cite{dompurify}.

\subsubsection{Audit query indistinguishability}

We empirically verify the indistinguishability of our proxy verification method via an experiment: We simulate an adversarial proxy's knowledge with a small dataset of 800 WildChat prompts representing normal user prompts, 100 audit prompts that the proxy knows are audit, and 100 audit prompts that the proxy does not know.
The proxy can obtain such a dataset simply by serving honestly for some time while collecting the queries.
The specific proportion of query types reflects the randomness of the audits, including our data poisoning approach.
We then fine-tune DeBERTaV3~\cite{he2023debertav3}, a state-of-the-art BERT-based model, on this dataset to predict whether a prompt is an audit.
We then test the model on another similarly-sized dataset with the same composition but different content.
The model only manages to achieve $0.5$-$0.6$ test AUC, which is not better than random guessing.
These results demonstrate the difficulty of distinguishing whether a query is an audit or not and the effectiveness of data poisoning.

\newpage
\section{Examples of Identity-Required Chatbots}

\begin{figure}[h!]
    \centering
    \includegraphics[width=0.9\linewidth]{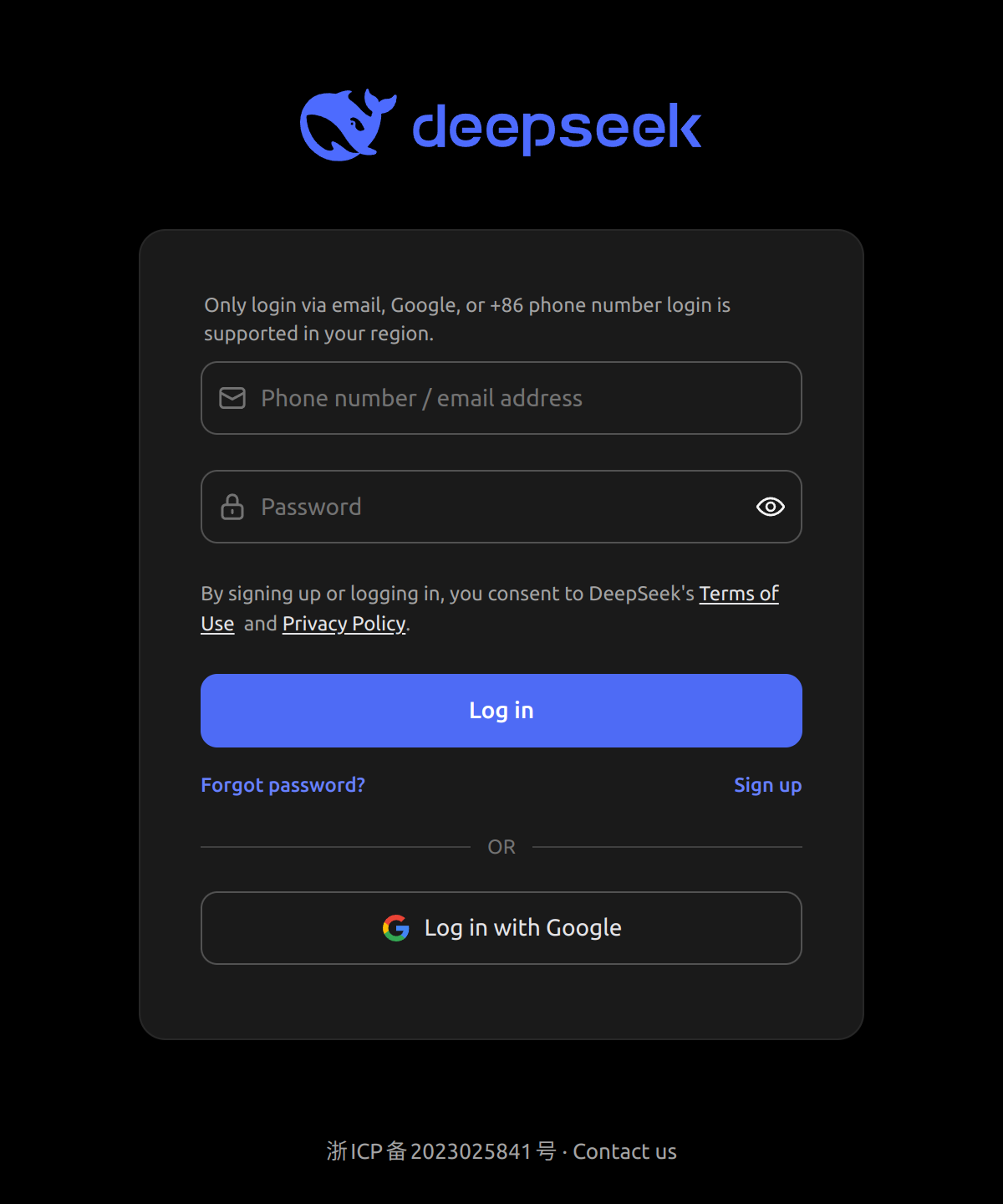}
    \caption{DeepSeek AI's landing page at chat.deepseek.com.}
    \label{fig:deepseek}
\end{figure}

\begin{figure}[h!]
    \centering
    \includegraphics[width=0.9\linewidth]{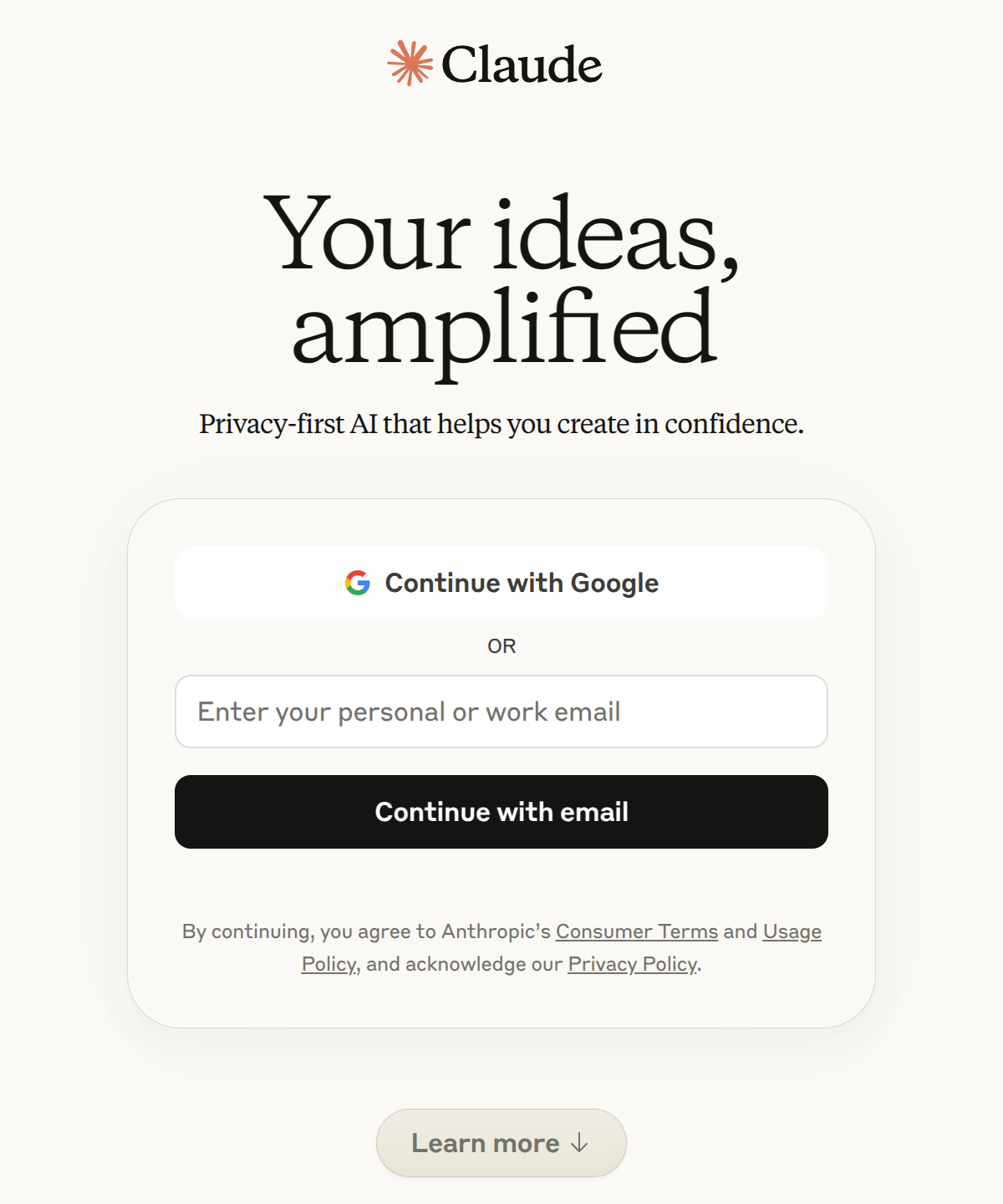}
    \caption{Anthropic Claude's landing page at claude.com.}
    \label{fig:claude}
\end{figure}

\begin{figure}[h!]
    \centering
    \includegraphics[width=0.9\linewidth]{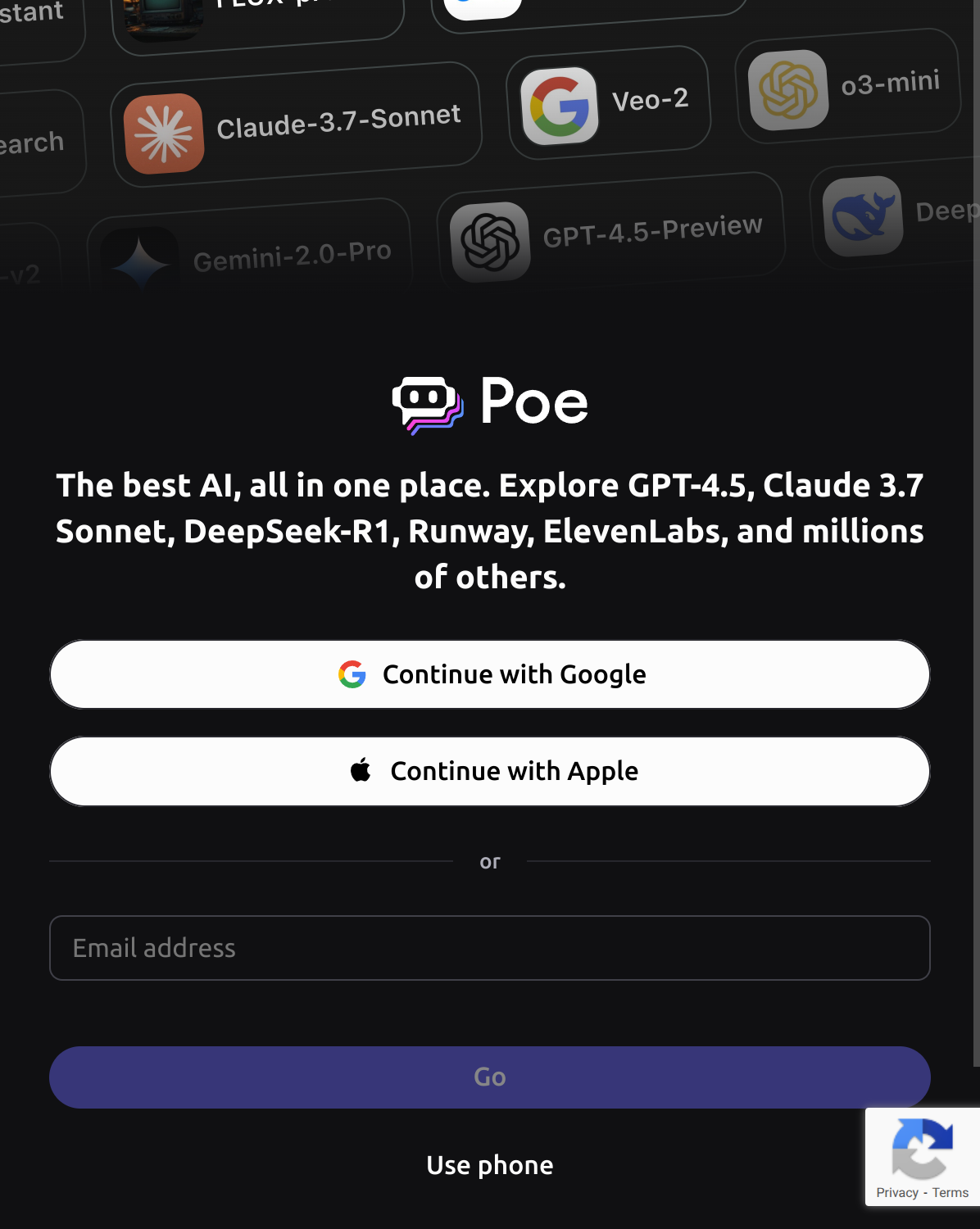}
    \caption{Quora Poe's landing page at poe.com.}
    \label{fig:poe}
\end{figure}

\end{document}